\RequirePackage{fix-cm}
\documentclass[twocolumn,epjc3]{svjour3}  
\smartqed  
\RequirePackage{graphicx}
\usepackage{graphicx}
\graphicspath{ {./figures/} }
\usepackage{siunitx}[=v2]
\usepackage{units}
\usepackage{listings}
\usepackage{color}
\usepackage{soul}
\usepackage[frozencache,cachedir=.]{minted}
\usepackage{amsmath}
\usepackage{amssymb}
\usepackage{xcolor}
\usepackage{hyphenat}
\usepackage{xcolor}
\definecolor{darkred}{rgb}{0.5,0,0}
\definecolor{darkblue}{rgb}{0,0,0.5}
\definecolor{firebrick}{rgb}{0.75,0.125,0.125}
\definecolor{darkgreen}{rgb}{0,0.5,0}
\usepackage[colorlinks=true,linkcolor=firebrick,citecolor=darkgreen,urlcolor=darkblue]{hyperref} 
\usepackage{lineno}
\usepackage{tikz}
\usetikzlibrary {angles,calc,quotes,intersections, decorations.pathmorphing}

\newcommand{\NuRadioMC}{\texttt{Nu\-Radio\-MC}}
\journalname{Eur. Phys. J. C}
\begin{document}
\title{NuRadioMC: Simulating the radio emission of neutrinos from interaction to detector}

\author{C.~Glaser\thanksref{e1,uci}
        \and
        D.~Garc\'{\i}a-Fern\'{a}ndez\thanksref{e2,desy,fau}
        A.~Nelles\thanksref{e3,desy,fau}
        \and
       J.~Alvarez-Mu\~{n}iz\thanksref{santiago}
       \and
       S.~W. Barwick\thanksref{uci}
       D.~Z.~Besson\thanksref{kansas}
       B.~A. Clark\thanksref{ohio}
       A.~Connolly\thanksref{ohio}
       C.~Deaconu\thanksref{chicago}
       K.~D.~de Vries\thanksref{vub}
       J.~C. Hanson\thanksref{whittier}
B.~Hokanson-Fasig\thanksref{madison}
R.~Lahmann\thanksref{fau,uci}
U.~Latif\thanksref{kansas}
S.~A.~Kleinfelder\thanksref{UCI2}
C.~Persichilli\thanksref{uci}
Y.~Pan\thanksref{delaware}
C.~Pfendner\thanksref{otterbein}
I.~Plaisier\thanksref{desy,fau}
D.~Seckel\thanksref{delaware}
J.~Torres\thanksref{ohio}
S.~Toscano\thanksref{ulb}
N.~van Eijndhoven\thanksref{vub}
A.~Vieregg\thanksref{chicago}
C.~Welling\thanksref{desy,fau}
T.~Winchen\thanksref{vub,mpifr}
S.~A.~Wissel\thanksref{calpoly}
 }
    
\thankstext{e1}{e-mail: christian.glaser@uci.edu}
\thankstext{e2}{e-mail: daniel.garcia@desy.de}
\thankstext{e3}{e-mail: anna.nelles@desy.de}


\institute{Department of Physics and Astronomy, University of California, Irvine, CA 92697, USA\label{uci}
           \and
          DESY, Platanenallee 6, 15738 Zeuthen, Germany \label{desy}
          \and
          Erlangen Centre for Astroparticle Physics, Friedrich-Alexander-Universit\"at Erlangen-N\"urnberg, 91058 Erlangen, Germany \label{fau}
          \and
          IGFAE \& Depto.~de F\'\i sica de Part\'\i culas, Universidade de Santiago de Compostela, Santiago de Compostela, Spain\label{santiago}
          \and
          Department of Physics and Astronomy, University of Kansas, Lawrence, USA \label{kansas}
          \and
          Dept.~of Physics and Center for Cosmology and AstroParticle Physics, The Ohio State Univ., Columbus, USA \label{ohio}
          \and
          Kavli Institute for Cosmological Physics, University of Chicago, USA\label{chicago}
          \and
          Vrije Universiteit Brussels, Belgium \label{vub}
          \and
          Whittier College Department of Physics, Whittier, CA, USA \label{whittier}
          \and
          Dept.~of Physics and Wisconsin IceCube Particle Astrophysics Center, University of Wisconsin, Madison, USA \label{madison}
          \and
          Department of Electrical Engineering and Computer Science, University of California, Irvine, CA 92697, USA \label{UCI2}
          \and
          Bartol  Research  Institute  and  Dept.  of  Physics  and  Astronomy,  University  of Delaware, Newark, USA \label{delaware}
          \and
          Otterbein University, Westerville, OH, USA \label{otterbein}
          \and
          Universit\'{e} Libre, Brussels, Belgium \label{ulb}
          \and
          Max-Planck Institute for Radio Astronomy, Bonn, Germany \label{mpifr}
          \and
          Physics Department, California Polytechnic State University, San Luis Obispo, CA, 93407, USA \label{calpoly}
}

\date{Received: date / Accepted: date}

\maketitle

    \begin{abstract}
   \NuRadioMC\ is a Monte Carlo framework designed to simulate ultra-high energy neutrino detectors that rely on the radio detection method. This method exploits the radio emission generated in the electromagnetic component of a particle shower following a neutrino interaction. \NuRadioMC\ simulates everything from the neutrino interaction in a medium, the subsequent Askaryan radio emission, the propagation of the radio signal to the detector and finally the detector response. \NuRadioMC\ is designed as a modern, modular Python-based framework, combining flexibility in detector design with user-friendliness. It includes a state-of-the-art event generator, an improved modelling of the radio emission, a revisited approach to signal propagation and increased flexibility and precision in the detector simulation. This paper focuses on the implemented physics processes and their implications for detector design. A variety of models and parameterizations for the radio emission of neutrino-induced showers are compared and reviewed. Comprehensive examples are used to discuss the capabilities of the code and different aspects of instrumental design decisions.
    \end{abstract}

\pagebreak

\section{Introduction}
High-energy neutrino astronomy is a most promising approach to address the still unanswered question of the origin of high-energy cosmic rays \cite{Ackermann:2019ows}. Neutrinos are the perfect messenger. Because they have negligible mass, are electrically neutral and have an extremely low interaction probability, they traverse the universe essentially unimpeded and point directly back to their sources. However, measuring neutrinos requires the instrumentation of large volumes to observe sufficient target material in which a rare interaction of these particles may occur. Currently the largest detector having observed neutrinos is IceCube, which uses the Antarctic ice as a target medium and instruments it with optical sensors \cite{IceCube}. 

Neutrino astronomy recently took a significant leap forward when the IceCube detector at the South Pole was used to measure a yet unexplained excess of events that provides the first strong evidence for astrophysical neutrino sources \cite{IceCube2015}. The sources have not yet been identified, though compelling evidence for a first source was delivered with the observation of a spatial and temporal coincidence between a flaring blazar, observed with gamma-ray telescopes, and a high-energy neutrino \cite{IceCube2018}. 
However, detection of astrophysical neutrinos above a few tens of PeV has not been achieved yet, possibly due to the neutrino flux expected to steeply fall with energy, which calls for instrumented volumes larger than those currently existing. A two orders of magnitude increase in the volume instrumented by IceCube is considered cost-prohibitive due to the attenuation and scattering of optical light in ice \cite{Aartsen:2013rt}. Such a detector may measure the continuation of the neutrino flux, as well as the expected fluxes in the ultra-high energy regime \cite{Ackermann:2019ows}.

\subsection{Experimental and physical context of radio detection}

High-energy neutrinos ($E_\nu > \SI{e16}{eV}$) can be most efficiently observed with the radio technique. Radio signals are produced via the Askaryan effect \cite{Askaryan} from particle cascades generated in the ice following interactions of the neutrinos. The Askaryan effect arises from the development of a charge excess in the shower front as it accumulates electrons from the surrounding medium. The resulting changing current leads to measurable radio emission in the MHz -- GHz frequency range. The Antarctic ice is transparent to these radio signals which allows for a cost-effective instrumentation of large volumes with sparse arrays. The attenuation length is about \SI{1}{km}, depending on the frequency and ice temperature \cite{barwick_besson_gorham_saltzberg_2005}. This results in an effective volume in the order of \SI{1}{km^3} per single detector station, similar to the size of the entire IceCube detector. 

The radio technique has already been successfully piloted with detectors at the South Pole and at Moore's Bay on the Ross ice-shelf. The ARIANNA project \cite{ARIANNA2015,ARIA} uses an array of autonomous detector stations with antennas located close to the ice surface, whereas the ARA project \cite{ARA} uses antennas at a depth of up to \SI{200}{m} below the firn layer. The experimental techniques matured substantially over the last years \cite{ARIANNAprogress,ARAprogress} and the community is well prepared for the construction of a large scale Askaryan detector with enough exposure to measure the continuation of the astrophysical neutrino flux to higher energies \cite{Ackermann:2019ows}, to potentially discover cosmogenic neutrinos \cite{1966PhRvL..16..748G,1966JETPL...4...78Z,1969PhLB...28..423B}, and measure particle physics properties at yet unachieved energies \cite{Ackermann:2019cxh}.

With the developments on the experimental side, improved Monte Carlo simulations became imperative, leading to the development of \NuRadioMC, which is presented in this article. A versatile and validated simulation of the radio signal in an Askaryan detector is crucial in many areas: for the determination of the sensitivity of a specific detector, for the optimization of the detector layout, to establish the requirements of the hardware to record the relevant parts of the signal, for the computation of a realistic signal expectation that is used to search for neutrino induced signals out of a large background of thermal and anthropogenic triggers, and finally, for the development of reconstruction techniques to determine the neutrino properties from the short radio flashes. In particular, the usage of modern deep-learning techniques requires a large and precise training data set. 

The diversity of possible station layouts (e.g.~compare the ARA and ARIANNA approach) requires a flexible software which is one of the main limitations of existing codes that were each targeted at a very specific experimental layout \cite{ThesisKamlesh,ARASim,Cremonesi:2019zzc}. \NuRadioMC\ is not tailored to a specific experimental design, and a detector station can have any number of antennas at arbitrary positions. In addition, the Askaryan radio technique is not limited to in-ice detectors. For example the lunar regolith has similar radio properties as ice and provides a immense neutrino target that can be observed from Earth with radio telescopes \cite{NuMoon,2017EPJWC.13504001J}, providing the opportunity for synergies in simulations. Hence, from the beginning \NuRadioMC\ was designed for maximum flexibility while maintaining user-friendliness. 

\subsection{Structure of \NuRadioMC}
The Monte-Carlo simulation of Askaryan signals from neutrino induced in-ice\footnote{We will continue to refer to the standard case of a neutrino interaction in ice, when describing NuRadioMC. However, the code is designed in such a way that it can also support media other than ice, and exotic particles such as for instance dark photons \cite{DarkPhotons}.} particle showers is logically split up into four independent steps, the four pillars of \NuRadioMC:
\begin{enumerate}
    \item \textbf{Event generation:} The simulation of a neutrino flux. This includes the simulation of different neutrino properties (energy, direction, flavor, etc.), lepton propagation, the position of the interaction vertices, and the properties of the induced particle sho\-wer, i.e., how much neutrino energy is transferred into the shower, whether it is an electromagnetic or hadronic shower, etc. 
    \item \textbf{Signal generation:} The calculation of the As\-ka\-ryan radio pulse generated from the particle shower.
    \item \textbf{Signal propagation:} The propagation of the radio signal through the medium, from its origin to each antenna. Naturally occurring media typically have a density gradient resulting in bent rather than straight trajectories of the radio signal. Also, multiple distinct paths from the interaction vertex to the antenna may exist for typical geometries and ice typically shows a frequency-dependent attenuation length. 
    \item \textbf{Detector simulation:} The simulation of all components of the detector hardware. This step includes the conversion from the electric-field pulses at the antenna positions to the measured voltages of each antenna channel, as well as the simulation of the trigger. It accounts for frequency dependent gain and group-delay, sampling-speed, record-length, etc.
\end{enumerate}

The separation of the four steps follows the temporal structure of the physical processes. In a MC simulation this sequence will be different and not linear, e.g., we determine the signal path before generating it, so that we only need to calculate the Askaryan signal at the particular emission angle leading to that path. Moreover, after having calculated the signal, we need to use the propagation module again to determine the signal attenuation along the path.

We note that the separation of signal generation and propagation is a valid approximation when the difference in travel time from different points of the emission region to an observer in a homogeneous medium and one in a medium with a density gradient (bent trajectories) is small with respect to the observation frequency. We find that this assumption holds for all but rare and extreme geometries of an in-ice detector at frequencies up to \SI{1}{GHz}.

The four pillars are complemented by a set of utility classes that are accessible at all times throughout the simulation such as a model of the medium, or a model of the signal attenuation. To ensure maximum flexibility and ease of use of different codes and programming languages the four pillars are separated as much as possible. The modules can be written in any language but Python wrappers of the relevant functions are required (this can be achieved e.g. with Cython \cite{Cython}), so that the simulation can be steered from Python. This design was chosen to maximize user-friendliness and allow for the interfacing with other existing frameworks.

\subsection{Improvements on the simulated physics in \NuRadioMC}
\NuRadioMC\ does not only improve in flexibility and ease of use over existing codes, but also includes more physics processes in the simulation than previous codes and improves on precision. In the event generation, the subsequent decay of taus following a tau-neutrino interaction is modelled and the interface to simulate any \emph{multi-bang} model is provided. Hence, models predicting several spatially-separated interactions can be implemented and simulated.

In the signal generation pillar, various Askaryan signal generation models are implemented. Previous MC codes relied on parameterizations of the frequency spectrum of radio emission \cite{AlvarezMuiz2000} or on time-domain calculations mostly restricted to electromagnetic shower profiles \cite{ARZ}. \NuRadioMC\ improves this approach by providing a time-domain calculation from an extensive library of electromagnetic, hadronic and tau-initiated showers. In particular, this allows for a realistic treatment of the Landau-Pomeranchuk-Migdal effect (LPM effect) \cite{Landau:1953um,Migdal:1956tc}.

In the signal propagation pillar, new ray-tracing techniques based on an analytic solution of possible signal paths are implemented. This implementation results in unprecedented combination of speed and accuracy. Furthermore, we provide the interface to a more detailed numerical calculation that can simulate the signal paths in arbitrary 3D density profiles.

In the detector simulation pillar, we use the \emph{NuRadioReco} code \cite{NuRadioReco} that allows for the simulation of any detector geometry. In particular, it includes a detailed antenna response for a variety of antenna types and arbitrary orientations, treating the full set of complex gains as well as complex triggers such as phased-arrays. 

In this article, we first describe each of the four pillars in detail and discuss different approaches. Then, we present three examples of how to use \NuRadioMC\ and discuss the implications for the design of a high-energy neutrino radio detector.


\section{Event generation}
\label{sec:event_generation}
The event generation is logically separated from the simulation and provides general event parameters as input to the simulation. The results of the event generation are stored in an HDF5 file \cite{hdf5}, which ensures that the event generator is easy to change in order to cover a variety of physics cases, as well as practical cases such as the simulation of calibration pulser data. This section describes the standard case implemented in \NuRadioMC\ and provides an outlook for future implementation and special cases. 

Having the event generation separated from the other simulation steps is beneficial because it allows the user to test the influence of different parameters on the same events. For example, the influence of different signal generation models, ice properties that influence the signal propagation or attenuation, and trigger schemes and thresholds, while using the same set of events. 

\begin{figure*}
    \centering
    \includegraphics[width=0.6\textwidth]{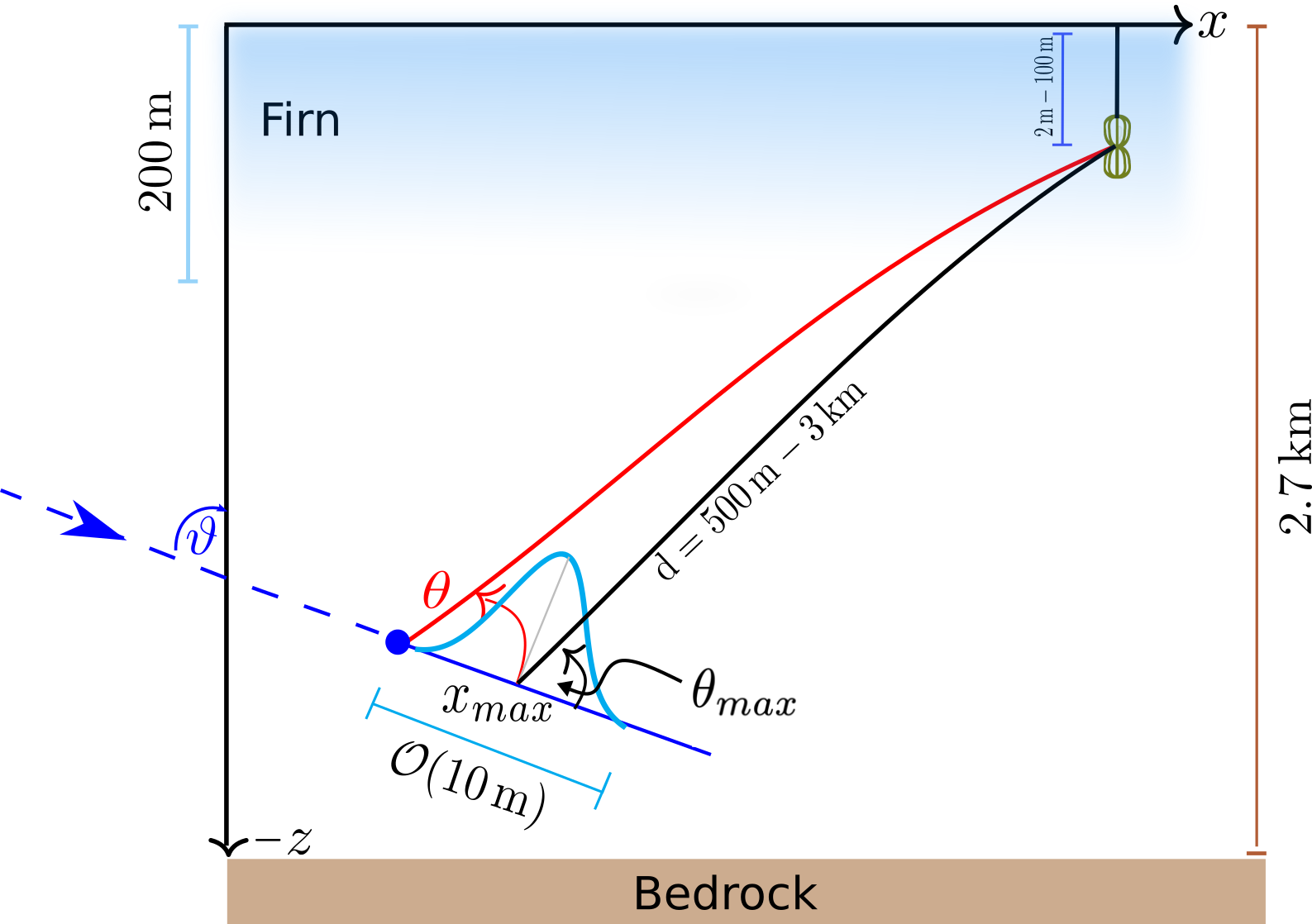}
    \caption{Sketch of the coordinate system used by \NuRadioMC\ and typical dimensions in the radio detection of neutrino interactions. The coordinate origin is at the ice surface. A quantity of particular interest is the viewing angle $\theta$, i.e., the angle at which the in-ice shower is observed. Due to the longitudinal extent of the shower, the viewing angle is not uniquely defined. By default, we measure the angle with respect to the neutrino interaction vertex, but sometimes it is appropriate to measure the angle with respect to the maximum of the charge-excess profile, which we denote with $\theta_\mathrm{Xmax}$. It should be noted that this is just one typical set-up, other choices of geometry are supported. }
    \label{fig:CS}
\end{figure*}

\subsection{Considerations concerning the coordinate system}

All coordinates are specified in a local Cartesian coordinate system with its origin centered at the surface of the ice (see Fig.~\ref{fig:CS}). The implementation of a global coordinate system that takes into account the curvature of the Earth is not required at this stage of precision: Due to attenuation of radio signals in the ice, the maximum propagation distance of radio signals is $\mathcal{O}(1-5)$~km where the impact of Earth attenuation is less than \SI{2}{m}. Thus, effects of Earth curvature can be ignored from the signal propagation step onwards. The maximum propagation distance also defines the necessary volume where neutrino interactions are simulated in. Thus, also for the standard event generation, a flat Cartesian coordinate system is sufficient. 

Earth curvature starts to matter in the tracking of tau leptons and simulation of their subsequent decay as the tau decay length can reach values above \SI{10}{km}. At \SI{10}{km} distance, the difference between a flat and curved surface is \SI{8}{m} which still small compared to the thickness of the ice sheet at the South Pole of \SI{2.7}{km}. Hence, the difference in target volume is also small. Another effect is that the probability of a neutrino reaching the simulation volume (referred to as \emph{neutrino event weight}, see Sec.~\ref{sec:earthattenuation}) is calculated based on the angle between the incident neutrino direction and the (flat) surface. Consequently, the neutrinos originating close to the horizon will have a systematic uncertainty in their assigned weights. However, at \SI{10}{km} distance, this effect is again small with a displacement of only \SI{0.1}{\degree}. In the future, effects of Earth curvature can be considered by correcting this angle in the neutrino event weight calculation. The additional complexity of implementing a global coordinate system does not seem required at this point. 

\subsection{Default event generator and file format details}

The default event generator creates a list of neutrino interaction vertices, specifies all relevant neutrino properties, and stores everything in an HDF5 file (see structure in \ref{sec:event_files}).

\begin{figure}
    \centering
    \includegraphics[height=0.25\textwidth]{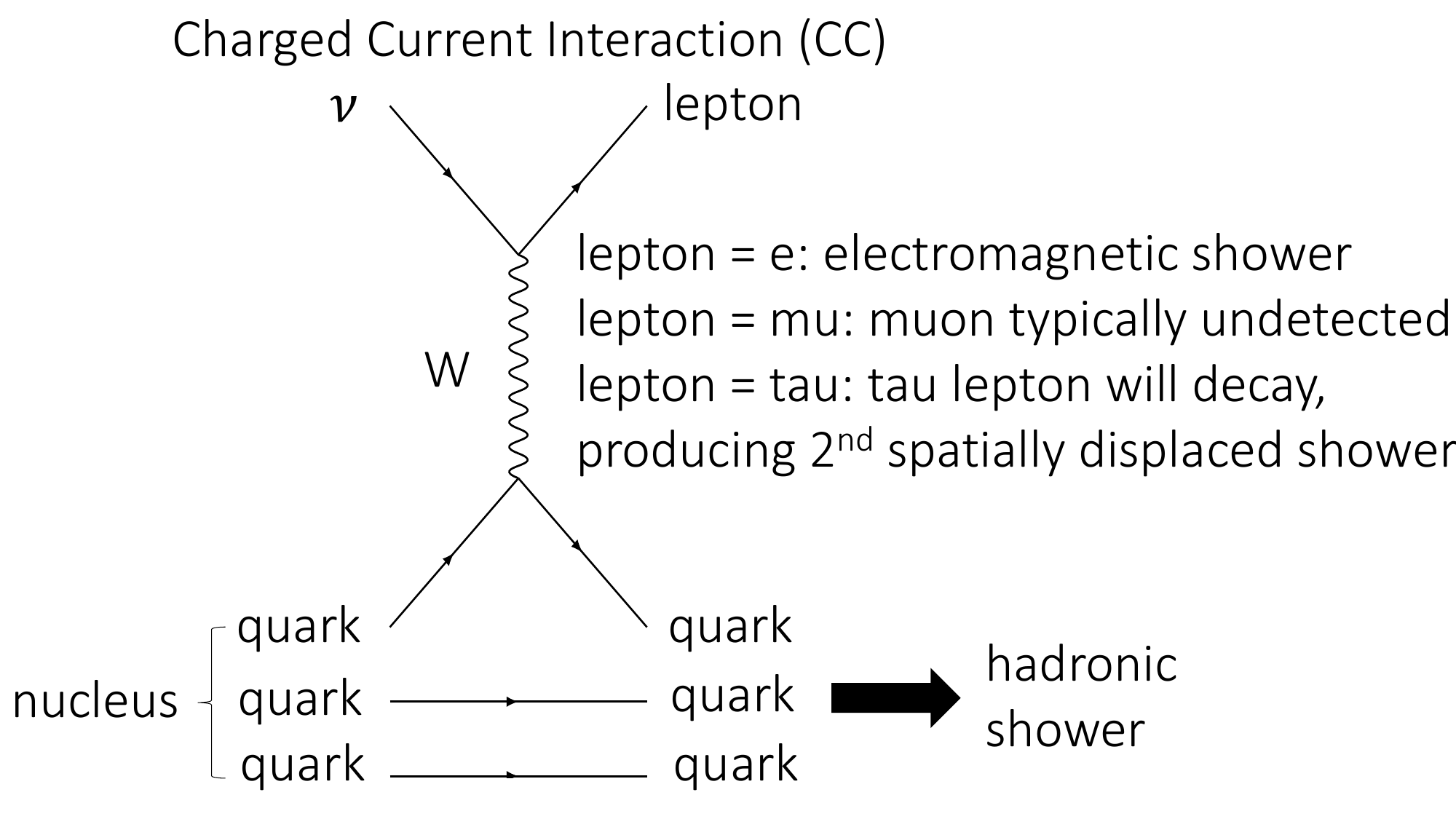}
    \includegraphics[height=0.25\textwidth]{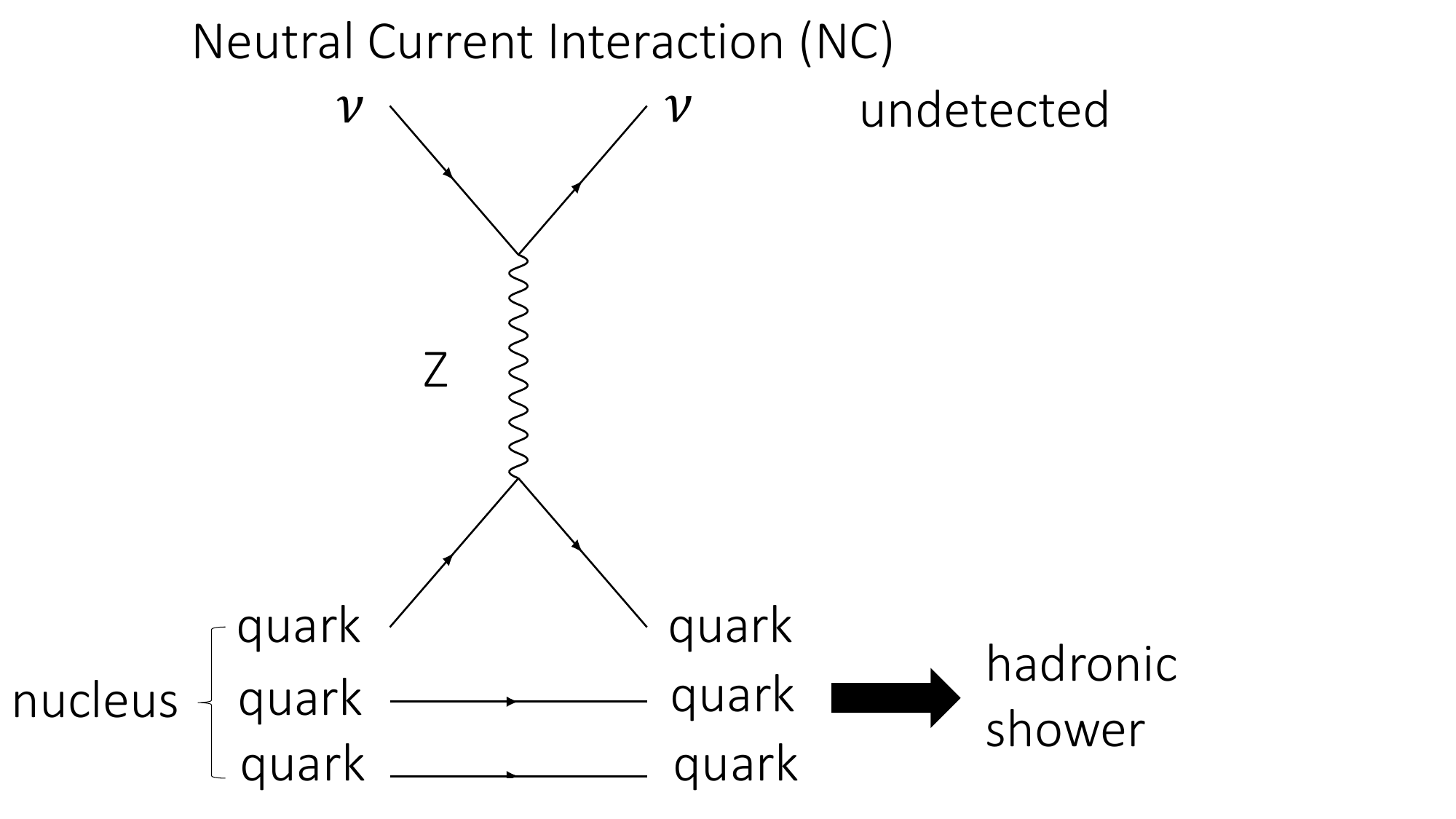}
    \caption{Feynman diagrams of a charged current and neutral current neutrino interaction.}
    \label{fig:ccnc}
\end{figure}

The event generator specifies the following parameters:
\begin{itemize}
\item the position of the neutrino interaction, randomly placed in a cylindrical volume surrounding the detector. The user can control the minimum and maximum radius and the vertical extent. 
\item the neutrino energy, drawn from a user definable energy spectrum between a minimal and maximal energy. We also allow to specify the \emph{deposited} energy instead, i.e., the amount of  neutrino energy that ends up in a particle shower producing an Askaryan signal.
\item the neutrino flavor. By default all flavors and par\-ticle\-/anti-particle nature have equal probability. Internally, this is specified using the Particle Data Group ID (PDGID) \cite{PDG}, which allows for cross-referencing with other Monte-Carlo codes. 
\item the neutrino direction. By default the full sky is uniformly covered but the user can restrict neutrino directions to specific ranges in zenith and azimuth angles. 
\item whether the neutrino undergoes a neutral current (NC) or charged current (CC) interaction (see Fig.~\ref{fig:ccnc} for an illustration of the two interaction types). We use a constant ratio CC:NC 0.7064:0.2936 according to the CTEQ4-DIS cross sections for the neutrino energy between \SI{e16}{eV} and \SI{e21}{eV} \cite{Gandhi1998}.
\item the inelasticity, i.e., the fraction of the neutrino energy going into the hadronic part of the interaction. The inelasticity distributions from \cite{Gandhi1996}, \cite{Connolly2011} and \cite{CooperSarkar:2011pa} have been implemented. 
\end{itemize}

We note that we place neutrino vertices with equal probability per volume. The probability of a neutrino reaching the detection volume is taken into account later by assigning a \emph{weight} to each event (see Sec.~\ref{sec:earthattenuation} for how the neutrino absorption is calculated). Similarly, it is currently ignored, if the density of the simulation volume is not uniform which changes the neutrino interaction cross section and thereby the interaction probability. As the density of the typical use-case of ice, only changes in the upper $\sim100$ m this effect is ignored at this stage of precision. It can be taken into account in the future by an additional weighting factor or by an event-by-event calculation of the neutrino cross section.

All these parameters are saved in a HDF5 table. This has several advantages. The data is saved efficiently, the format is platform and programming-language independent, stand-alone viewers exist to quickly inspect the files, and apart from storing the actual data tables, it allows saving meta attributes such as the parameters the event set was generated for. 

Typical data sets consist of millions of events which would take too long to simulate in a single process. Therefore, the event generator allows to automatically split up the data set into smaller chunks, i.e., into separate HDF5 files with typically 10,000 to 100,000 events per file. Then, the \NuRadioMC\ simulation can be performed for each file separately, and we provide the tools to merge the individual output files back together. 

\subsection{Multiple showers}
Previous radio simulations only considered particle showers created by the initial neutrino interaction. However, in case of charged current interactions of muon and tau neutrinos, the produced muons and taus might interact or decay producing a second spatially displaced particle shower that generates Askaryan radiation. 

The typical decay length of a tau lepton range from \SI{50}{m} at tau energies of \SI{1}{PeV} to \SI{50}{km} at tau energies of \SI{1}{EeV}. This increases the sensitivity of an Askaryan detector because tau neutrinos can interact far away from the detector but still produce a visible signal if the tau happens to decay close enough to the detector. 

Muons in turn are unlikely to decay but they can undergo a catastrophic $dE/dX$ energy loss, depositing a substantial fraction of their energy into the ice and initializing a hadronic shower \cite{Koehne:2013gpa,Dunsch:2018nsc}. 
In general, more exotic models can also be considered that predict multiple spatially displaced showers per neutrino. Hence, \NuRadioMC\ offers the flexibility to specify an arbitrary number of interaction vertices per event. This is incorporated into the file format by inserting additional events into the event list with the same event ID. 

We consider several levels of detail. While a simple treatment of tau decays exists in \NuRadioMC\ itself, we also foresee the inclusion of more complete particle decay codes, such as PROPOSAL \cite{Koehne:2013gpa,Dunsch:2018nsc} that tracks secondary losses of all types of lepton. 

\begin{figure*}
    \centering
    \includegraphics[width=0.7\textwidth]{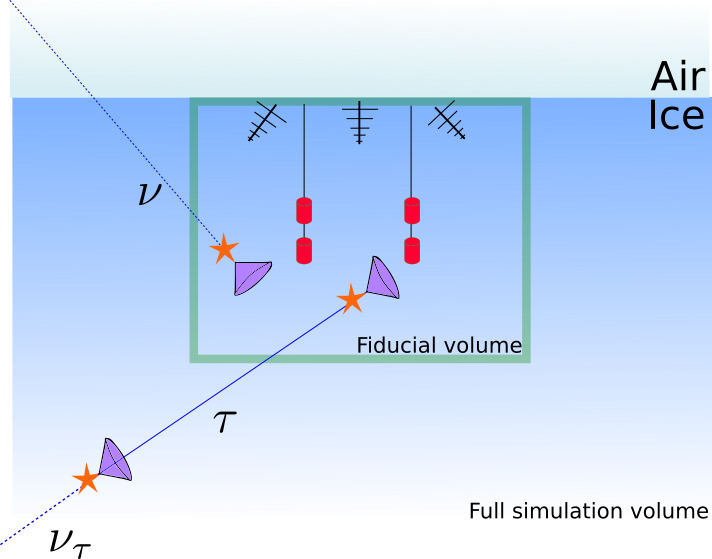}
    \caption{Sketch of the geometry and the concept of a fiducial volume of the event generator. Neutrinos tracks are generated in a full simulation volume, but only the radio emission of primary or secondary interactions are considered, when they take place in a fiducial volume encompassing the detector.}
    \label{fig:fiducial}
\end{figure*}

\subsection{Tau neutrinos}
\label{sec:tau}
In \NuRadioMC, for the first time in an in-ice simulation, we provide the inclusion of secondary sub-showers from tau-decays that add additional detection channels, flavor sensitivity and contribute to the effective volume.

Due to the large decay length of tau leptons, a large volume needs to be simulated to catch the few cases in which there is a secondary interaction close enough to the detector. This increases the computation time enormously as it scales proportionally to the simulated volume, and makes this brute-force approach unfeasible. Therefore, we developed the following technique: we generate neutrino interactions in an arbitrarily large volume including all secondary interaction vertices (e.g. from tau decays) but save only those primary and secondary interactions that take place in a much smaller fiducial volume surrounding the detector while keeping track of the total number of simulated events (see Fig.~\ref{fig:fiducial} for an illustration). The user needs to make sure that the fiducial volume is chosen large enough such that the probability to trigger the detector is negligible for interaction vertices outside of this volume. This allows for a computationally efficient simulation of complex physics models. 

Once a tau is created after the interaction of a tau neutrino in the volume, we calculate its decay time $t_{\mathrm decay}$ and energy at decay. We first randomly sample a decay time $\tau_{\mathrm decay}$ in the tau particle rest frame from an exponential distribution using a mean tau decay lifetime $2.903 \times 10^{-13}$ s \cite{PhysRevD.98.030001}. If the tau energy is less than $E_\tau = $\SI{1}{PeV}, we do not account for tau energy losses along the path, and the decay time is simply given by the product of the Lorentz factor $\gamma$ and the sampled decay time $\tau_{\mathrm{decay}}$ in the tau rest frame
\begin{equation}
    t_{\mathrm{decay}} = \gamma(E_\tau) \tau_{\mathrm{decay}}.
\end{equation}
The decay length $l_\tau$ is calculated multiplying $t_{\mathrm{decay}}$ by the particle speed, while the energy of the $\tau$ at decay is equal to the initial tau energy.

\begin{figure}
    \centering
    \includegraphics[width=0.49\textwidth]{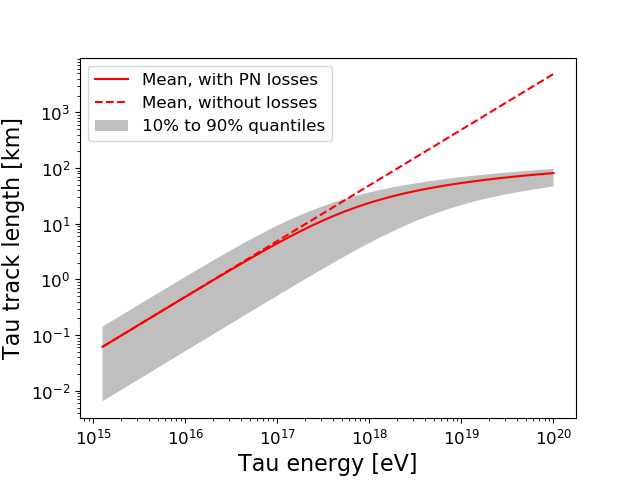}
    \includegraphics[width=0.49\textwidth]{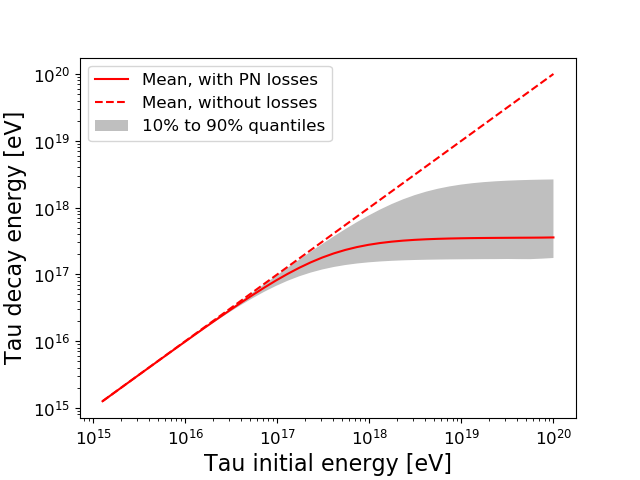}
    \caption{Top: Tau decay length as a function of the tau energy. 
    Bottom: Tau decay energy as a function of the initial tau energy.
    Due to the one-tailed nature of
    the exponential decay function, we show the decay length for the mean proper decay time with photonuclear losses (solid line) and without any losses (dashed line).
    The shaded band represents the area spanning from the $10\%$ proper decay time quantile to the $90\%$ quantile ($80\%$ of total probability). This implementation matches what has been shown previously in \cite{PhysRevD.63.094020}.}
    \label{fig:taulength}
\end{figure}

In the case the tau has an energy greater than \SI{1}{PeV}, we include photonuclear tau energy losses in our calculation.
These are not very well constrained and we use a simple model inspired by
the results in \cite{taulosses}. We take the mean energy loss per amount of traversed matter in ice to be,
\begin{equation}
    \left\langle \frac{\mathrm{d}E_\tau}{\mathrm{d}X} \right\rangle \approx f(E_\tau) =
    b_1 E_\tau + b_2 E_\tau \log_{10}(E_\tau/E_0),
\label{eq:csda}
\end{equation}
with $b_1=\SI{1e-7}{cm^2/g}$, $b_2=\SI{1.8e-7}{cm^2/g}$, and $E_0=\SI{1}{PeV}$. Above $E_\tau=E_0$,
it is a good approximation to assume that the tau speed is equal to the speed of light in vacuum $c$. This allows us to write
the time $t$ that it takes a tau with initial energy $E_{\tau,i}$ to reach a lower energy $E_\tau$ as,
\begin{equation}
    t(E_\tau) = \frac{1}{c \rho_\mathrm{ice}} \int_{E_{\tau,i}}^{E_\tau} \frac{\mathrm{d}{E'}}{f({E'})}.
\end{equation}
Once $t(E_\tau)$ is known, we numerically obtain the inverse function $E_\tau(t)$ for equally-spaced
times by interpolation. The decay time is obtained by solving the following integral equation for $t_{\mathrm{decay}}$:
\begin{equation}
    \int_0^{t_{\mathrm{decay}}}  \ \frac{m_\tau}{E_\tau(t)} \mathrm{d}t = \tau_{\mathrm{decay}},
\end{equation}
from which the tau decay length above \SI{1}{PeV} is obtained as:
\begin{equation}
    l_\tau \approx c t_{\mathrm{decay}}.
\end{equation}
In Fig.~\ref{fig:taulength}, left, we show the decay length $l_\tau$ as a function of tau energy. The straight dashed line
represents the mean decay length without tau energy losses, which increases linearly with energy. The solid line indicates the decay length assuming that the
decay time in the rest frame is equal to the mean decay time $\tau_{\mathrm{decay}}$ and accounting for deterministic tau-energy losses during propagation given in Eq.~\eqref{eq:csda}. The shaded band represents an $80\%$ confidence interval for the decay length, where the decay time has been drawn from an exponential distribution. Stochastic energy losses have not been accounted for.
In Fig.~\ref{fig:taulength}, right, we show the tau energy at decay obtained with the same assumptions used for obtaining the tau decay length shown in the left panel.
Tau energy losses become important around \SI{100}{PeV}.

\subsection{Options for additional physics processes or calibration purposes}
The event generation described above is the default event generator in \NuRadioMC. However, emission from a standard-model neutrino-induced shower is only one possible scenario that can be covered. The users have the freedom to implement their own event generators according to other physics assumptions, e.g., new physics or for simulating calibration signal generators. We provide an example to simulate a calibration measurement online \cite{Example4}. As long as the events are saved according to the well-defined file structure, \NuRadioMC\ can process any input files. A skeleton event generator is provided as an example \cite{EvtGenSkeleton}.


\section{Signal generation}
\NuRadioMC\ provides several modules for the generation of the radio signal from showers. The user may choose from a selection ranging from well-known frequency-domain parameterizations of the Askaryan signal to a state-of-the art semi-analytic calculation. 

A uniform interface in the form of a simple function is provided for all models (see \cite{Link:online:documentation} and List.~\ref{lst:signalgeneration} in \ref{sec:implementation}). In this way the \NuRadioMC\ code also serves as a reference implementation for all models. Furthermore, the well-defined interface allows for an easy extension of \NuRadioMC\ with additional models. Even calibration emitters can be (and are) implemented to simulate a calibration measurement with \NuRadioMC.

In the following, we first present the different signal generation models available in \NuRadioMC\ before discussing their differences and giving recommendations for use in different cases. We discuss a variety of models, some for more pedagogical reasons, others because they are fast, and others because they are accurate. We hope that this section also serves as reference discussion of several widely used emission models, however, it is not an attempt at completeness.


\subsection{Frequency-domain parametrizations}

\NuRadioMC\ currently provides two frequency-domain parameterizations of the Askaryan signal. One, referred to as \emph{Alvarez2000}, is also used in the simulation code for the ANITA detector (IceMC) \cite{Cremonesi:2019zzc} and for the ARIANNA array (ShelfMC) \cite{PersichilliThesis,ThesisKamlesh}, and is an implementation of the parameterization of \cite{AlvarezMuiz2000}, which was validated against a full simulation of Askaryan radiation performed with the ZHS Monte Carlo \cite{ZHS}. This is a microscopic simulation of the shower and its radio emission, that does not contain signal propagation and detector simulation.

The other parameterization (\emph{Alvarez2009}) is an updated version of the first one.
It is based on the so-called ``box model'' of shower development \cite{Alvarez_box} and separate parameterizations for electromagnetic \cite{Alvarez2012} and hadronic \cite{Alvarez2009} showers are provided. Both parameterizations are the product of three functions. The first is a scaling function $A$ that grows linearly with the primary energy $E_0$, frequency $f$, and the sine of the observing angle $\theta$.
The second and third functions are two continuous cutoff frequency factors $d_L$ and $d_R$ that account for deviations from linearity due to incoherence effects associated to the longitudinal and lateral extensions of the shower. For electromagnetic showers, the LPM effect is modelled including random fluctuations of the size of the effect. 


Although we encourage the use of the \emph{Alvarez2009} parameterization, we have also included the older parameterization \emph{Alvarez2000} for comparison with previous work and other codes. The latter can be understood as a simplified version of the former, with constant factors, a simple continuous cutoff factor instead of two, and a Gaussian function for the dependence of emission on viewing angle. Because of its simplicity, it provides qualitative and easily understandable, however, not necessarily precise insights into the main dependencies of the Askaryan signal. For pedagogical reasons, we explicitly provide the parameterization of this model here and give an example of the resulting Askaryan signals. 

If the shower is observed on the Cherenkov angle, the electric field (scaled to a distance of \SI{1}{m}) according to \emph{Alvarez2000} is given by
\begin{equation}
  \frac{\varepsilon_{c}^{1\mathrm{m}}}{\unit{V/m/MHz}}(E_{sh},f) = 2.53 \times 10^{-7} \cdot \frac{E_{sh}}{\text{TeV}} \cdot \frac{f}{f_0} \cdot \frac{1}{1 + (\frac{f}{f_0})^{1.44}} \, ,
  \label{eqn:SelfMCEFieldAmp}
\end{equation}
with the shower energy $E_{sh}$, frequency $f$ and $f_0 = \unit[1.15]{GHz}$.
Signal amplitudes off the Cherenkov cone, $\varepsilon^{1\mathrm{m}}$, are modeled as a Gaussian profile according to 
\begin{multline}
  \varepsilon^{1\mathrm{m}}(E_{sh},f,\theta_v) =  \varepsilon_{c}^{1m}(E_{sh},f) \cdot \\ \frac{\sin \theta_v}{\sin \theta_c} \cdot \exp\bigg[-\ln 2 \cdot \Big(\frac{\theta_v - \theta_c}{\sigma_\theta}\Big)^2\bigg]
  \label{eqn:ShelfMCConeAngle}
\end{multline}
with $\varepsilon_{c}^{1m}$ given in Eq.~\eqref{eqn:SelfMCEFieldAmp}, and where $\theta_v$ is the viewing angle relative to the shower axis. The angular width of the cone around the Cherenkov angle $\sigma_\theta$ is a function of both frequency and energy. For hadronic showers $\sigma_\theta$ is given in Eq.~(6) of \cite{AlvarezMuiz1998}, for which a factor to account for the so-called missing energy, energy going mainly into muons and neutrinos that does not contribute to the Askaryan signal, is included in Eq.~\eqref{eqn:SelfMCEFieldAmp}.

\begin{figure*}
  \begin{centering}
    \includegraphics[width=\textwidth]{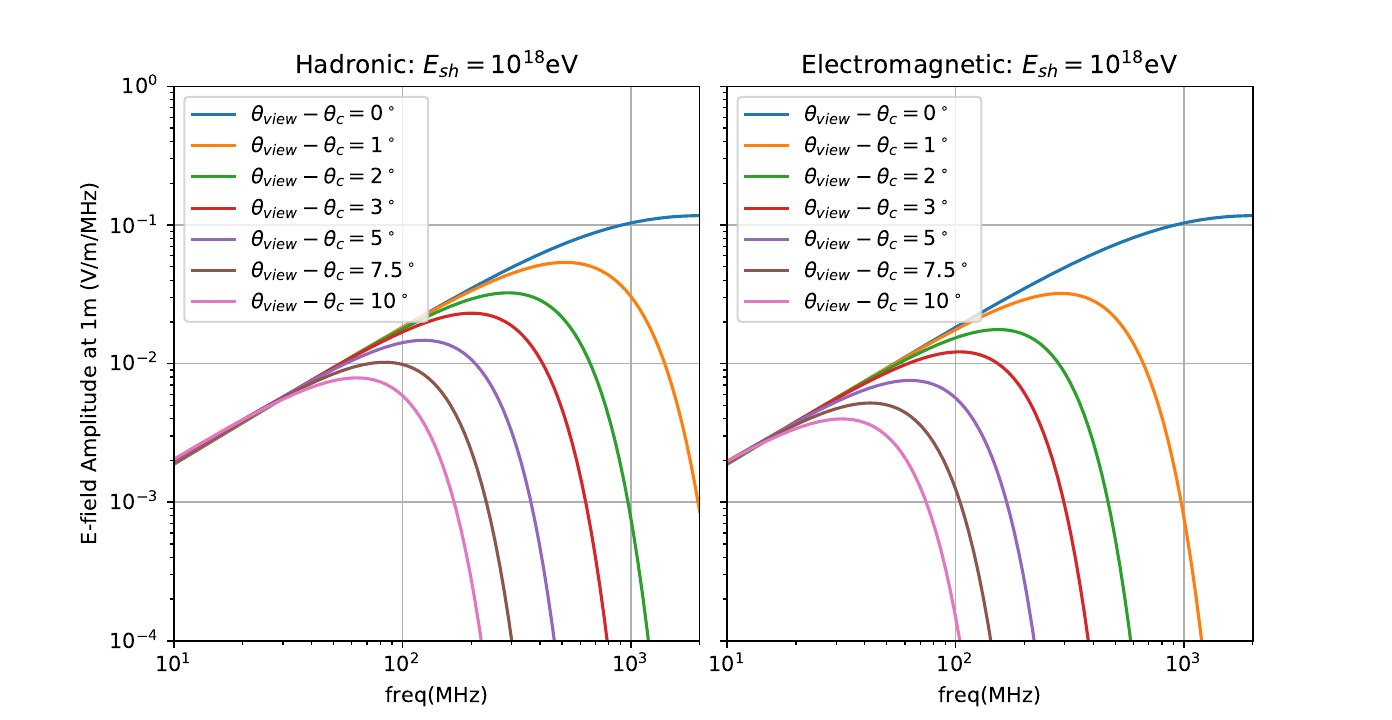}
    \caption{Electric field amplitude $\varepsilon^{1m}$, $\unit[1]{m}$ from the neutrino interaction vertex (Eq.~\eqref{eqn:ShelfMCConeAngle}) for hadronic (left) and electromagnetic (right) showers with $E_{sh} = \unit[10^{18}]{eV}$ using the parameterization \emph{Alvarez2000}.  Note that as the viewing angle shifts away from the Cherenkov cone angle, high frequency components fall off. For the EM showers, the cone width $\sigma_\theta$ is reduced due to the LPM effect.}
    \label{fig:ShelfMCEField1mEMHad}
  \end{centering}
\end{figure*}

For electromagnetic showers above \SI{2}{PeV}, the shower profile becomes elongated due to the Lan\-dau-Po\-meran\-chuck-Migdal (LPM) effect. In the simple model of \emph{Alvarez2000} such an elongation corresponds to a reduced $\sigma_\theta$ which is modeled according to the prescription in \cite{AlvarezMuiz1997}.  This in turn manifests itself as a rapid decrease in the high frequency content of the Askaryan signal off the Cherenkov cone for EM showers, as seen in Fig.~\ref{fig:ShelfMCEField1mEMHad}.

For \NuRadioMC, the time-domain signal based on \emph{Alvarez2000} and \emph{Alvarexz2009} is generated by taking the simple approximation of a phase that is constant with frequency and equal to $90^\circ$, yielding a bipolar pulse in the time domain.


\subsection{Fully analytic treatment including the LPM effect and Cascade Form Factor}

\NuRadioMC\ provides an implementation of the analytic model of Askaryan radiation (\emph{HCRB2017}) \cite{Hanson2017} that builds on previous work by \cite{RB}. This fully analytic model accounts simultaneously for the three-dimensional \textit{form factor} of the cascade, and the cascade elongation.  The form factor is the spatial Fourier transform of the instantaneous charge distribution of the cascade. The form factor affects the Askaryan signal properties in the same way a multi-pole filter affects any time-domain signal. Although some authors have provided partial solutions for the three-dimensional form-factor in the past \cite{HU2012421}, in \cite{Hanson2017} a complete solution is presented that includes dependence on the viewing angle $\theta$.  This allows for the analytic exploration of the relevant parameter space affecting $\sigma_{\theta}$ and $\sigma_{\nu}$, the width of the Cherenkov cone and the Fourier spectrum, respectively.

This module builds upon the work of \cite{RB} where the authors provide analytic functions for Askaryan radiation correct in both the near and far-field regimes.  When a cascade is elongated due to the LPM effect, both regimes become important given the three-di\-men\-sional nature of the form-factor. \emph{HCRB2017} treats the LPM effect as a smooth stretching of the shower profile using the results of \cite{Gerhardt:2010bj}.

The fully analytic nature of this model has the advantage that it gives direct insights into the physical dependencies of the Askaryan signal. However, as shown in the radio emission of air showers \cite{EVA2012} a purely analytic model comes at the cost of a poorer accuracy.


\subsection{Semi-analytic model in the time domain}
\begin{figure*}
    \centering
    \includegraphics[width=0.7\textwidth]{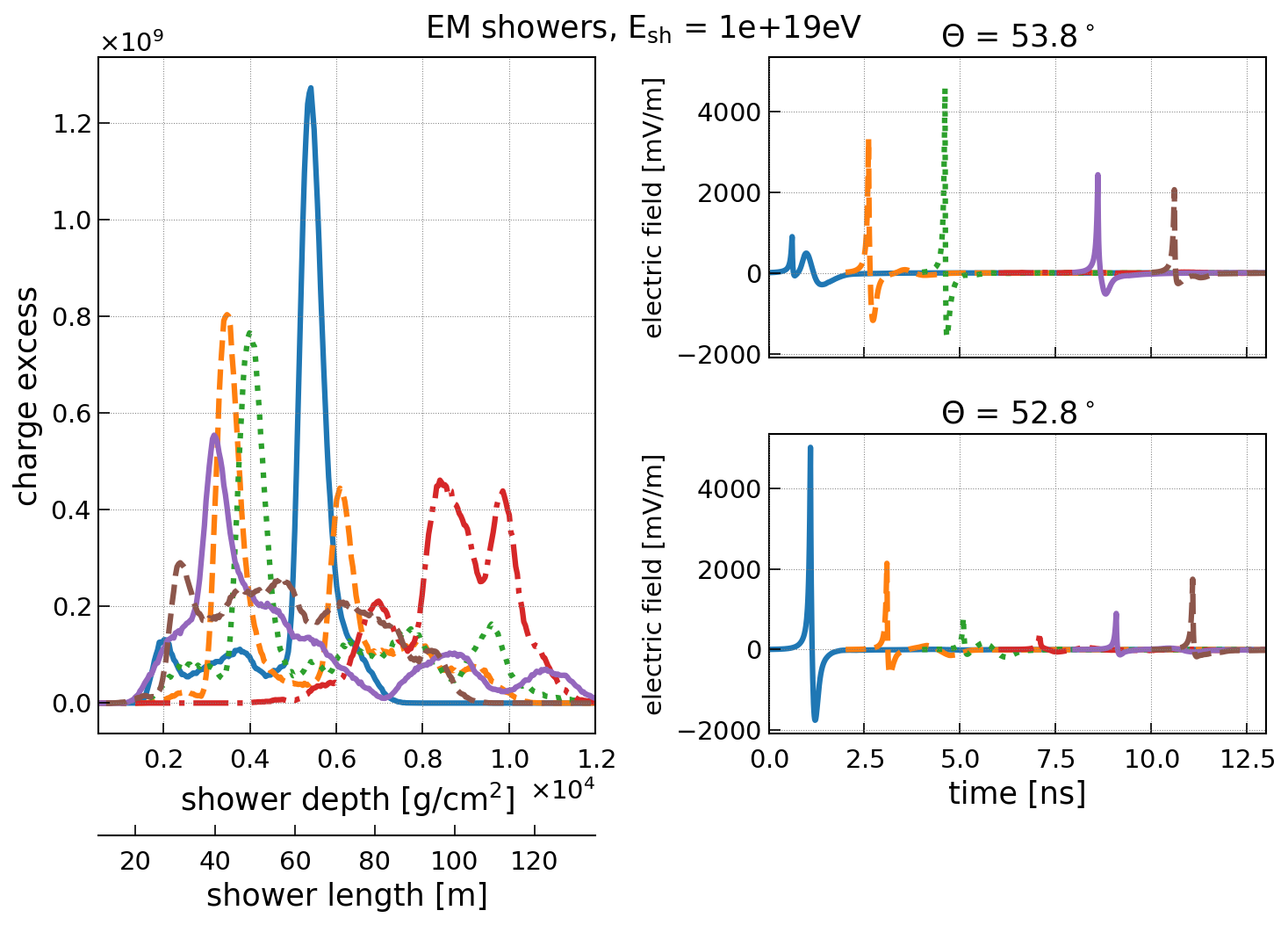}
    \caption{Charge-excess profiles and resulting Askaryan signal (unfiltered). (left) Charge-excess, i.e., number of electrons minus number of positrons, as a function of shower depth and length of six electromagnetic shower with an initial energy of \SI{e19}{eV}. The variation in the charge-excess profile is due to the stochastic nature of the shower development effected by the LPM elongation. (right) The resulting Askaryan signal for the charge-excess profiles according to the \emph{ARZ} model for two different viewing angles at \SI{1}{km} distance. The pulse start time is shifted for a better visibility of all pulses.}
    \label{fig:ARZ_EM19}
\end{figure*}

A third option for the signal generation is to calculate the Askaryan radiation individually from detailed charge-excess profiles in the time domain, following the approach in \cite{ARZ2}. The implementation in \NuRadioMC\, referred to as \emph{ARZ}, is based on a realistic shower library. This allows to precisely model the effects of LPM elongation \cite{Landau:1953um,Migdal:1956tc} and the resulting large shower-to-shower fluctuations on the Askaryan signal on a single event basis, rather than describing an average behaviour. The model also captures subtle features of the cascades like sub-showers and accounts for stochastic fluctuations in the shower development which can alter the Askaryan signal amplitudes significantly (see e.g.~discussion in \cite{Alvarez2009} or Fig.~\ref{fig:ARZ_EM19}). This model is the most accurate treatment of Askaryan radiation implemented in \NuRadioMC, but it comes at the expense of larger computation times as it involves computationally expensive convolutions of the Askaryan vector-potential with Monte-Carlo generated cascade profiles.

The main idea behind the \emph{ARZ} method is that the electromagnetic vector potential $\mathbf{A}$ in Coulomb gauge can be expressed as an integral in shower depth containing the shower profile, a factor that accounts for polarization, another factor that accounts for distance to the emitting point of the shower, and a form factor $F_p$: 
\begin{multline}
    \mathbf{A}(r,z,t) = \frac{\mu}{4\pi} 
    \int_{-\infty}^{\infty} \mathrm{d}z'
    \frac{Q(z')}{\sqrt{r^2 + (z-z')^2}} \mathbf{p}(z') \\
    F_p \left( t - \frac{z'}{v} -
    \frac{n\sqrt{ r^2 + (z-z')^2} }{c} \right),
    \label{eq:ARZ}
\end{multline}
where $r$ is the radial distance of the observer to the shower, $z$ is the vertical coordinate of the observer, $z'$ is the shower depth, $Q(z')$ the excess charge profile, $\mathbf{p}$ is the polarization vector and $F_p$ is the form factor (see \cite{ARZ2} for more details).
This form factor $F_p$ has approximately the same shape for every particle shower in ice, which allows us to treat it as a constant function. It only depends on the type of the shower, i.e., hadronic or electromagnetic, and a parameterization of the form factor for both shower types is provided.

The charge profile $Q(z')$ depends on the nature of the shower (hadronic or electromagnetic), the shower energy, and is also subject to random fluctuations. The LPM effect, for instance, modifies the charge profile, which in turns modifies $\mathbf{A}$ through Eq.~\eqref{eq:ARZ}. All the physical processes that are relevant for the electric-field calculation contribute to $Q(z')$, so as long as a correct description of the charge profile is available in the shower library, an accurate electromagnetic potential $\mathbf{A}$ can be calculated with Eq.~\eqref{eq:ARZ}.

Once $\mathbf{A}$ is known, the radiation electric field can be calculated with a derivative, since in Coulomb gauge 
$\mathbf{E}_\mathrm{rad} = -\frac{\partial \mathbf{A}}{\partial t}$.
The agreement between the electric field predicted by the ZHS Monte Carlo and the one obtained with the \emph{ARZ} model is quite satisfactory, yielding a few percent of error up to \SI{2}{GHz} (see Fig.~3 in \cite{ARZ}). The \emph{ARZ} model considers that the shower has a volume and therefore is adequate for computing the fields of observers near the shower as long as the considered wavelengths are small with respect to the distance to the shower.

\NuRadioMC\ provides a modern Python-based implementation of the code used in \cite{ARZ2} and optimized routines for numerical integration. The code includes a shower library of charge-excess profiles for different sho\-wer types:
\begin{enumerate}
    \item electromagnetic: purely electromagnetic showers from $\nu_e$ charge current interactions.
    \item hadronic (neutrino): showers started by the fragmentation of the nucleon struck by the neutrino, i.e., the result of neutrino neutral current interactions and the hadronic part of an electron neutrino charged current interaction. 
    \item hadronic (tau): showers initiated by a hadronic decay of a tau lepton. A tau decay into muons will not produce any significant shower, and tau decays into electrons correspond to purely electromagnetic showers. 
\end{enumerate}
The last category is not simulated explicitly. Instead, the branching ratios of a tau decay and the fraction of energy ending up in the particle cascades is parameterized using the results of \cite{Koehne:2013gpa,Dunsch:2018nsc}. Then, the shower library of electromagnetic (category 1) or hadronic (category 2) showers is used with the appropriate shower energy. We note that the initial hadronic particles that start the hadronic shower are different between a fragmenting nucleon and a hadronic tau decay. This might lead to small differences in the hadronic shower developments. However, for now we ignore this subtle difference and use category 2 also for hadronic tau decays. In the future, we will provide a separate shower library for category 3. Currently, \NuRadioMC\ comes with version 1.2 of the shower library that will be described in the following. 

The showers were simulated using HERWIG \cite{Bellm:2017bvx} for the simulation of the first neutrino nucleon interaction, and ZHAireS \cite{AlvarezMuniz:2011bs} for the subsequent simulation of the particle shower in ice.
The charge-excess profiles are binned in bins of \SI{37}{g/cm^2} for electromagnetic showers and \SI{18}{g/cm^2} for hadronic showers. To optimize the computation speed, we integrate Eq.~\eqref{eq:ARZ} numerically using the trapezoid rule given the binning of the charge-excess profile. The form factor is a strongly peaked function which requires a more precise integration around the peak. This is achieved by dynamically interpolating the charge-excess profile at the positions corresponding to the peak of the form factor. 

The shower library (version 1.2) contains 10 showers for every shower energy ranging from  \SI{e15}{eV} to \SI{e20.5}{eV} in steps of $\Delta\log_{10}(E) = 0.1$ for both electromagnetic and hadronic showers. To obtain charge-excess profiles for shower energies that were not explicitly simulated we do the following:
At first order, the charge-excess amplitude scales with shower energy. Hence, in a simulation, we pick one shower realization randomly from the nearest energy bin and re-scale the charge-excess amplitude by $E_\mathrm{event}/E_\mathrm{library}$.

\begin{figure*}
    \centering
    \includegraphics[width=0.7\textwidth]{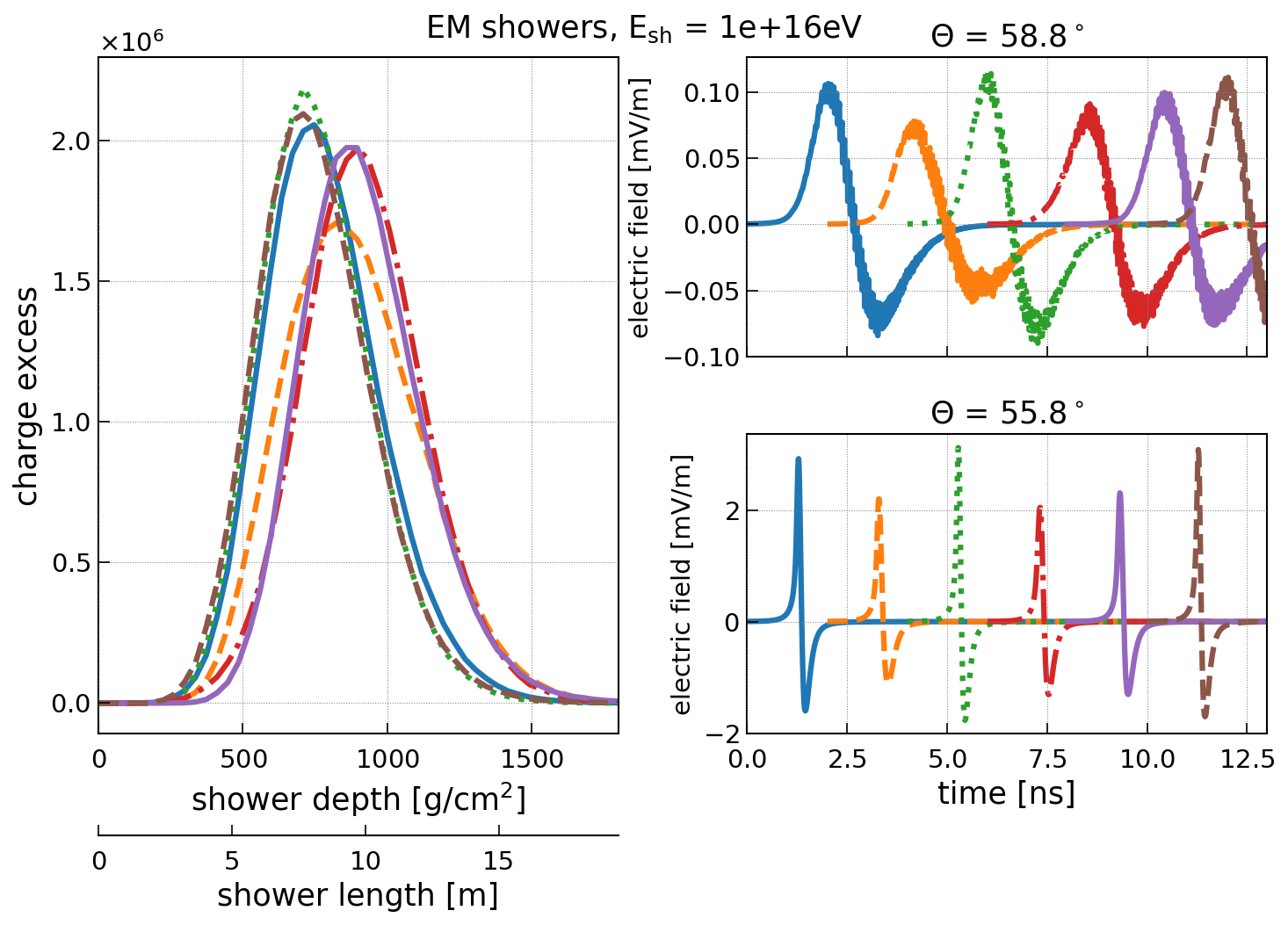}
    \caption{Charge-excess profiles and resulting Askaryan signal (unfiltered). Same as Fig.~\ref{fig:ARZ_EM19} but for electromagnetic showers with an initial energy of \SI{e16}{eV}. At this energy the LPM effect only has a small influence on the shower development and stochastic shower-to-shower fluctuations are small.}
    \label{fig:ARZ_EM16}
\end{figure*}

\begin{figure*}
    \centering
    \includegraphics[width=0.7\textwidth]{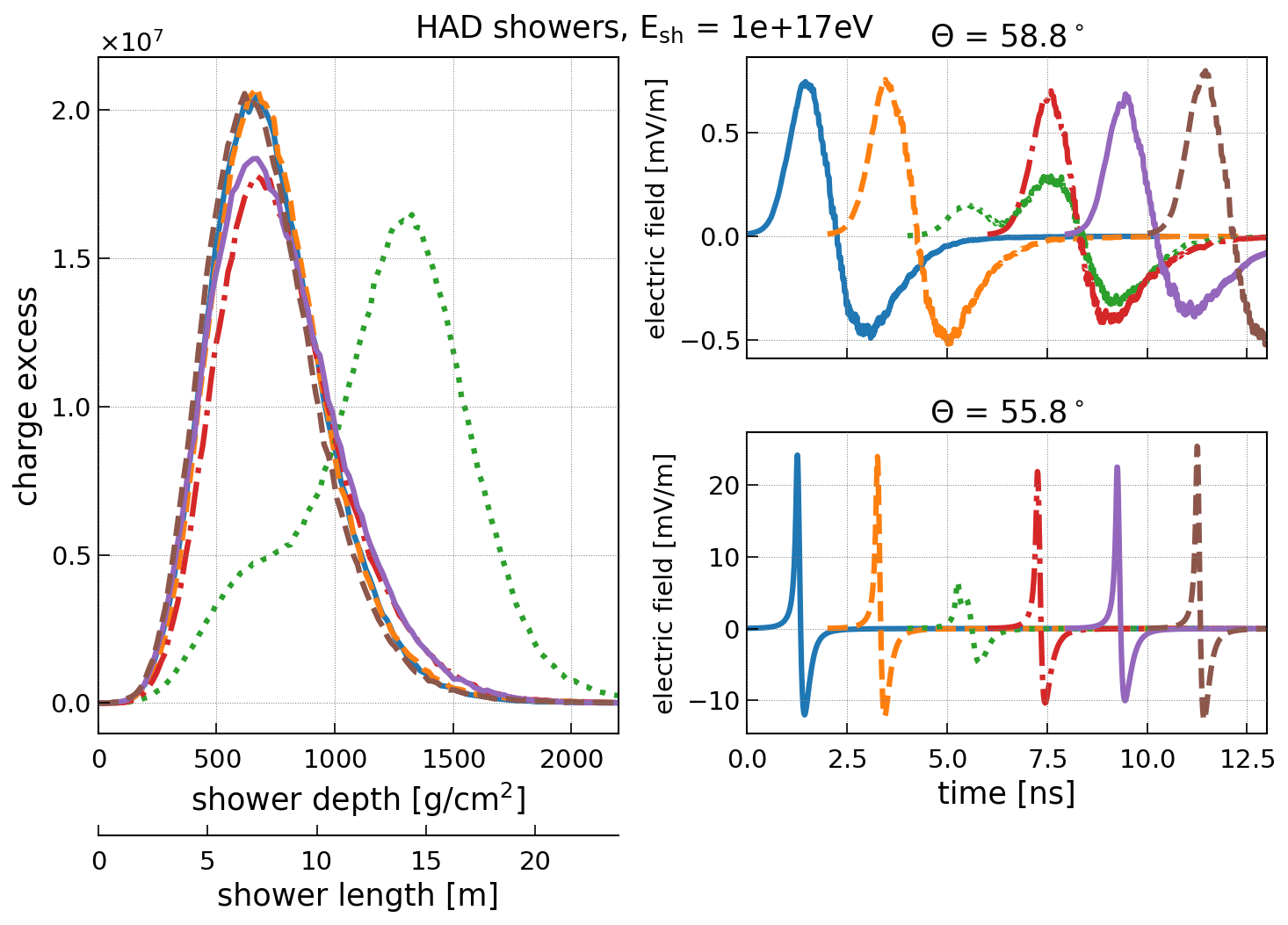}
    \caption{Charge-excess profiles and resulting Askaryan signal (unfiltered). Same as Fig.~\ref{fig:ARZ_EM19} but for hadronic showers with an initial energy of \SI{e17}{eV}. Most hadronic showers are not influenced by the LPM effect and show only very little shower-to-shower fluctuations. Different energies mostly scale the charge-excess and electric-field amplitudes approximately linear with energy but have a small effect on the shower length. However, sometimes a high-energy $\pi^0$ that is created in one of the first interactions decays instead of interacting leading to an electromagnetic sub-shower that experiences significant LPM elongation (green dotted curve in this figure).}
    \label{fig:ARZ_HAD17}
\end{figure*}

To discuss and illustrate the improvement in accuracy when using the \emph{ARZ} approach as opposed to a parameterization, we consider the influence of the LPM effect on the radio signal. The main consequence of the LPM effect is that the interaction probability of high-energy electrons, positrons and photons is suppressed leading to an elongation of the shower profile. The strength of the effect is proportional to the energy of the particle. Therefore, it mostly affects highly-energetic electromagnetic showers above a few PeV in ice, in which a large amount of energy is carried by individual particles. Previously in the literature (e.g. \cite{AlvarezMuniz:2011bs,Hanson2017}), the effect was often modelled via stretching of a smooth shower profile. However, this does not take into account the stochastic nature of the process and the fact that the first few particles of an electromagnetic shower are impacted differently by the LPM effect as the energy is not equally distributed. As a consequence, one gets multiple spatially displaced EM showers as shown in Fig.~\ref{fig:ARZ_EM19}. In this figure, also the resulting Askaryan signals are shown for two different viewing angles $\theta$ which are significantly different for different realizations of the shower (see Fig.~\ref{fig:CS} for a sketch of the coordinate system). 
Low energy EM showers are less influenced by the LPM effect and the resulting Askaryan signals are similar for all shower realizations (cf. Fig.~\ref{fig:ARZ_EM16}).
Hadronic showers exhibit little shower-to-shower fluctuations except for the rare cases where a high-energy electromagnetic shower is initiated in one of the first interactions that then gets LPM elongated (see Fig.~\ref{fig:ARZ_HAD17}).


\subsection{Comparison of models}

\begin{table*}
\begin{center}
\begin{tabular}{ p{3cm} p{5cm} p{5cm}  }
 
 model & advantages & shortcomings \\ 
 \hline\hline
parameterization  (\emph{Alvarez2009}) & fast, accurate representation of the signal amplitudes, includes statistical fluctuations from LPM & no phase information, only valid in far-field \\ 
 fully analytic (\emph{HCRB2017})& fast, phase information provided, valid in near and far-field, LPM is treated as elongated shower & no statistical fluctuations from LPM, generalization, absolute amplitudes less accurate   \\ 
 semi analytic (\emph{ARZ}) & phase information provided, near and far-field, realistic LPM treatment based on simulated shower library  & computationally expensive\\
 full MC & precise modelling of all details of shower development & slow, no implementation in \NuRadioMC\ yet\\
\end{tabular}
\end{center}
\label{tab:Askaryan}
\caption{Overview of alternative methods implemented in \NuRadioMC\ to calculate the signal following a neutrino interaction}
\end{table*}

\begin{figure*}[t]
    \centering
    \includegraphics[width=0.49\textwidth]{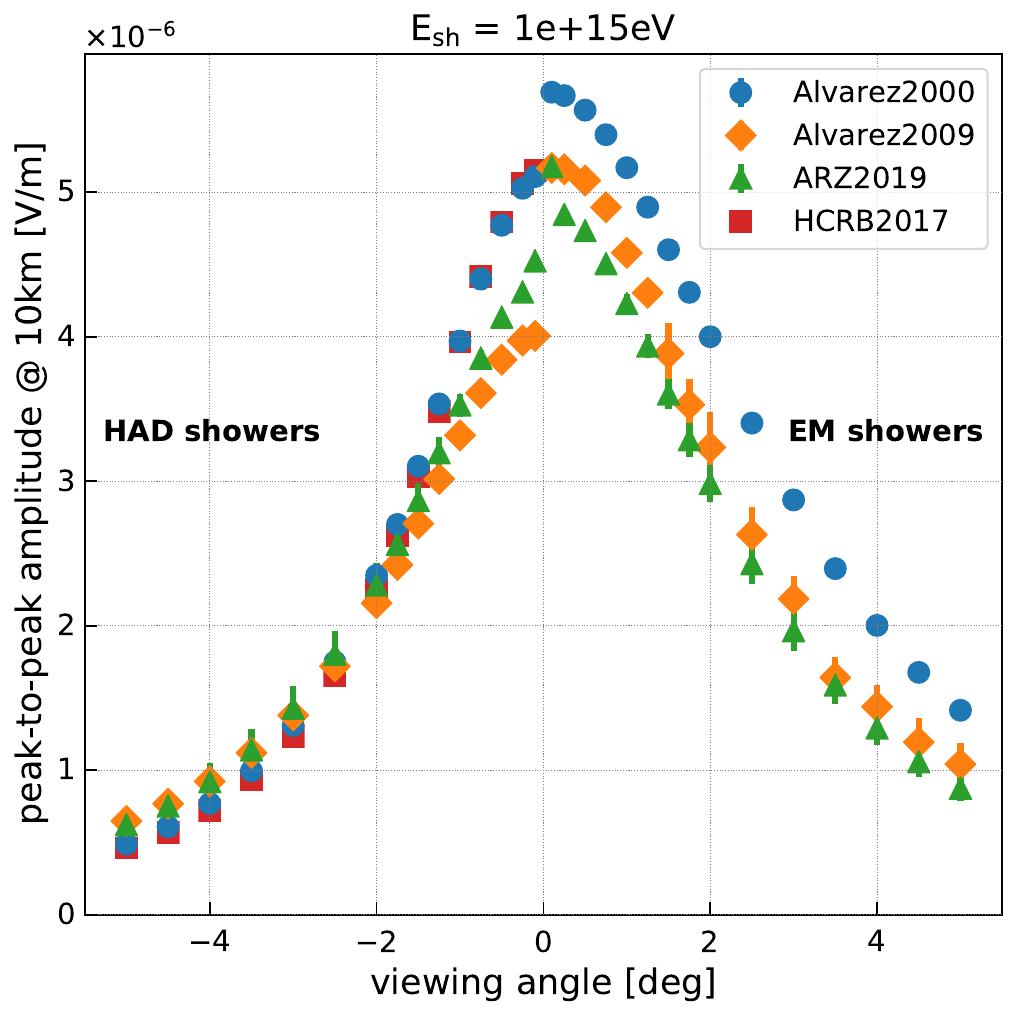}
    \includegraphics[width=0.49\textwidth]{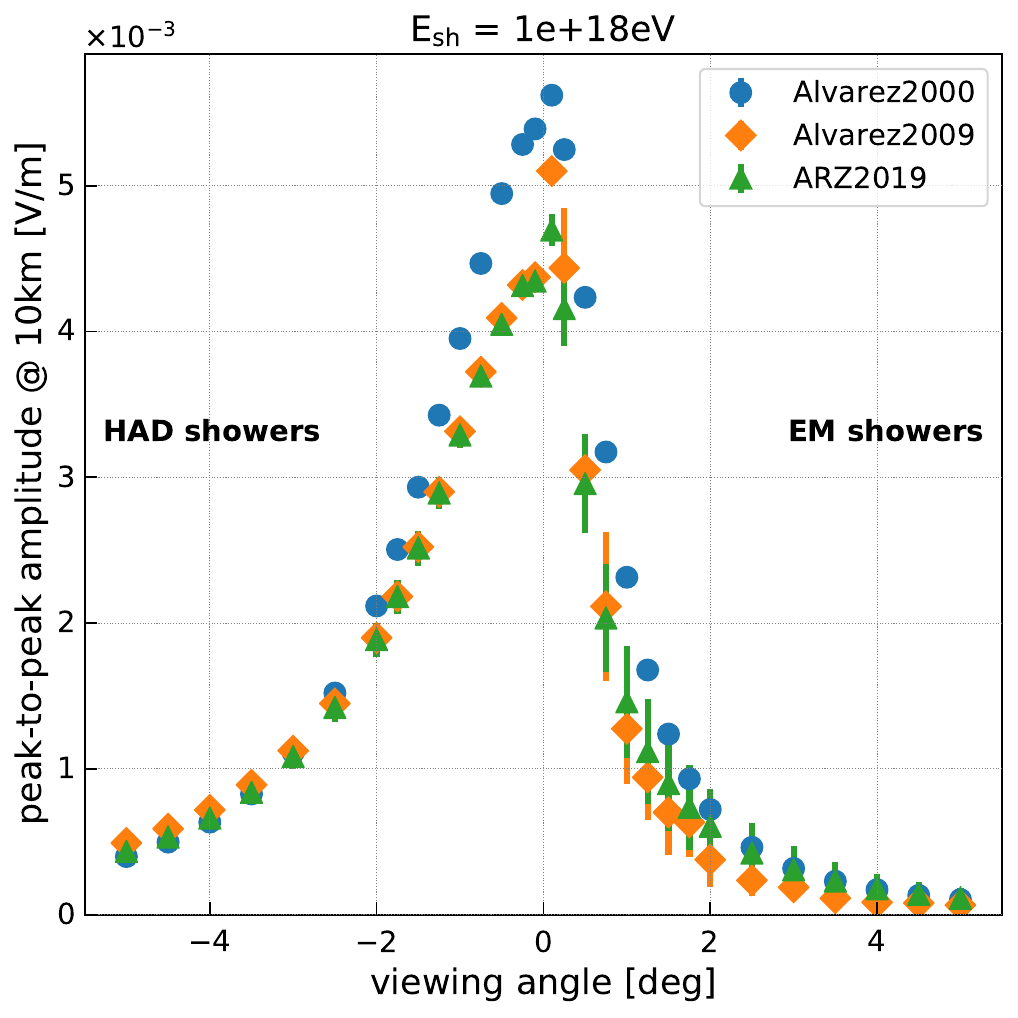}
    \caption{Comparison of Askaryan models. Shown the peak-to-peak amplitude (the difference between the maximum and the minimum of the Askaryan pulse) as a function of viewing angle. The pulses are filtered in a typical experimental bandwidth of \SI{100}{MHz} to \SI{500}{MHz}. The left part of the plot (negative angles) shows the prediction for hadronic showers and the right part of the plot (positive angles) the prediction for electromagnetic showers of the same shower energy. (left) \SI{e15}{eV} shower energy. (right) \SI{e18}{eV} shower energy.}
    \label{fig:ask_comparison}
\end{figure*}

Each signal generation module in \NuRadioMC\ has its own strengths and shortcomings. We first compare the signal models with respect to their resulting signal properties and then discuss practical considerations.
We provide a quick overview of the discussion in Tab.~\ref{tab:Askaryan}.  In Fig.~\ref{fig:ask_comparison}, a comparison of the predicted peak-to-peak amplitudes in a typical detector bandwidth of \SI{100}{MHz} - \SI{500}{MHz} is presented that will be discussed below.

The frequency-domain parameterizations are based on a detailed full Monte Carlo simulation of the particle shower and a calculation of the resulting radio signal using the ZHAireS code \cite{Alvarez2012}. Thus, their predictions of the signal amplitudes are accurate, the narrowing of the Cherenkov cone due to the LPM effect is modelled and even statistical fluctuations in the shower development are parameterized (only \emph{Alvarez2009}). The models are fast to evaluate and the computing time is negligible compared to the other parts of the simulation. We also provide an older version, \emph{Alvarez2000}, that was most commonly used in previous simulation frameworks and is therefore important for comparison. However, we strongly recommend the usage of the newer model \emph{Alvarez2009} as the older model typically overestimates the Askaryan amplitudes by roughly 20-30\%. 
The \emph{Alvarez2009} model is in good agreement with the more precise \emph{ARZ} time-domain calculation (cf. Fig.~\ref{fig:ask_comparison}). 

The main shortcomings of such parametrizations are that no phase information is provided which leads to inaccuracies in the time domain. Typically, the phases are approximated as constant \SI{90}{\degree} as function of frequency, which results in a perfectly symmetric bipolar pulse. While this may be a reasonable approximation for many cases, it does not capture the details of the shape of the pulses and does not account for physical time delays. Thus, these models are suitable for general sensitivity calculations given the correct prediction of amplitudes. However, more detailed models are recommended to study trigger efficiencies and event reconstruction that are based on pulse shape and timing. 

Another option is the fully analytic model \emph{HCRB2017} that also calculates the phases and is thus suitable for the time-domain. It provides helpful insights into the dependencies of the Askaryan signals on shower elongation and shower width. As being analytically it does not model the statistical fluctuations occurring in showers that can be substantial as shown in Fig.~\ref{fig:ARZ_EM19}. The signal strength prediction depends strongly on the longitudinal cascade width $a$, which has to be approximated with a Gaussian function for different cases (electromagnetic, hadronic and LPM showers). The approximations lead to a mis-match between the predictions of this model and the ones of the other models that are based on a microscopic Monte Carlo simulation where the calculation of the radio signal is based on first principles resulting in a few percent accuracy as shown in the radio emission of air showers \cite{Gottowik_2018}. In particular, the \emph{HCRB2017} model overpredicts the amplitudes at higher shower energies and the reduction of the cone width due to the LPM effect. Therefore, we only show the \emph{HCRB2017} model for low-energy hadronic showers in Fig.~\ref{fig:ask_comparison}. Furthermore, the treatment of pulse arrival times is complex in an analytic model, complicating the integration with the different signal propagation modules (see Sec.~\ref{sec:propogation}). Naturally, the model is computationally very fast given its analytic approach. 

The semi-analytic model \emph{ARZ} builds on a shower library of charge-excess profiles and thus models all details like sub-showers including statistical fluctuations in the shower development. The calculation is performed in the time domain. It therefore includes all phase information and gives an accurate prediction of the pulse shape and timing. The model provides valid results even when the distance from observer to shower is comparable to or smaller than the shower dimensions, as long as the distance is large compared the considered wavelengths. Above \SI{100}{MHz}, and at distances greater than \SI{10}{m}, the use of the ZHS formula, on which the ARZ model is based, is justified \cite{zhs-valid}. It is the most precise model available and recommended for the development of neutrino identification and reconstruction algorithms. Its disadvantage is that it is computationally more expensive. In a full end-to-end simulation it takes up roughly 90\% of the computing time. When using this model, the computing time increases roughly by a factor of 10. 

The next level of precision can be achieved with full Monte Carlo simulations where each shower particle is tracked and the radio emission is calculated from the acceleration and creation of each charged particle. This is done for air showers in codes like CoREAS \cite{Coreas} and ZHAireS \cite{AlvarezMuniz:2011bs}, which are required to achieve the necessary accuracy for modern air shower experiments that are pushing the reconstruction uncertainties (e.g.~\cite{AERAPRL,GlaserErad2016,Gottowik_2018,LOFARCr}). Currently, there is no urgency to require this level of accuracy for neutrino predictions, given the experimental uncertainties and the computational costs of a full Monte Carlo. However, future developments like a next generation of CORSIKA \cite{Engel:2018akg} are followed closely to allow for synergies and compatibility with \NuRadioMC.  

One could also consider another future improvement in the combination of signal generation and propagation. As discussed earlier, the decoupling of signal generation and propagation leads to noticeable inaccuracies in an inhomogeneous medium (where the signal trajectories are bent, cf. next section) if the extent of the emission region becomes large with respect to the distance to the receiver and if the trajectory is substantially refracted in the firn. Then, the time delay of the propagation time from different emission points to the receiver vary between a homogeneous and inhomogeneous medium, so that signal generation and propagation cannot be separated without loss of accuracy. This effect can be taken into account naturally in a microscopic Monte Carlo simulation by calculating the (curved) path from each emission point to the observer. In an intermediate step, one could use the ARZ2019 model, where the Askaryan signal is calculated from the charge-excess profile to address the issue: Instead of calculating the emission from the full charge-excess profile at once, a shower can be subdivided into small chunks. The Askaryan radiation can then be calculated per chunk and propagated individually to the receiver.


\section{Signal propagation}
\label{sec:propogation}
The signal propagation pillar of \NuRadioMC\ handles the propagation of the Askaryan signal through the medium to the observer positions. Like the other pillars, this part of the code is clearly separated so that different signal propagation modules can be implemented and exchanged by the user. This is achieved by defining an interface in form of a Python class (see general example in \cite{base:Signal:prog:class}).

\begin{figure*}
    \centering
    \includegraphics[width=0.7\textwidth]{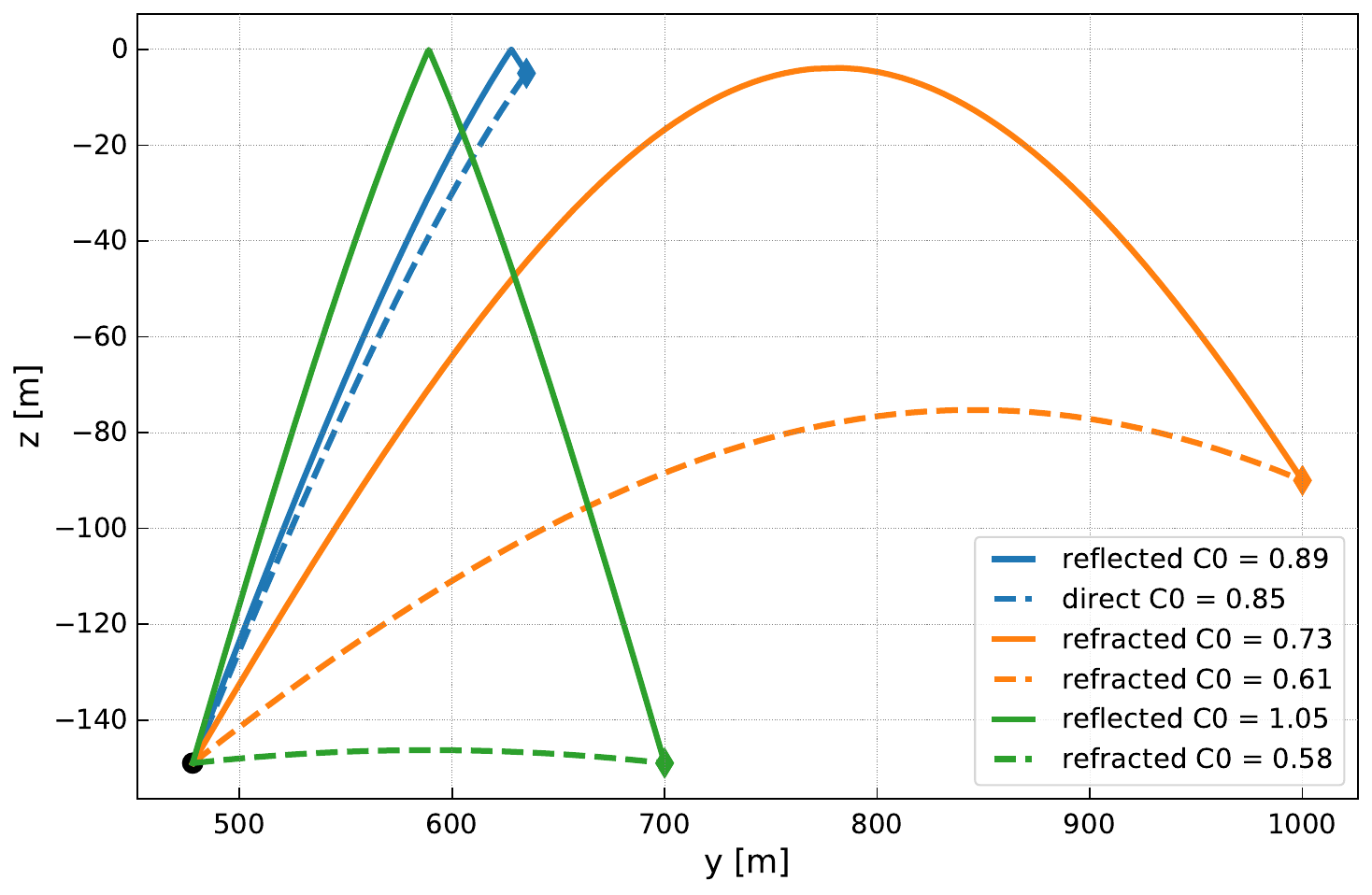}
    \caption{Example of typical ray-tracing solutions for receiver locations differing in depth and horizontal distance to a given emitter. The emitter is indicated by the black circle at the bottom left. Lines of the same color belong to the same receiver location. Shown are the combinations of direct and reflected ray (blue), refracted and reflected ray (green), and two refracted rays (orange). The numbers in the legend show the $C_0$ parameter of Eq.~\eqref{eq:analytic_solution} that defines the shape of the curve.}
    \label{fig:raytracing}
\end{figure*}

The signal propagation problem is typically approximated via ray tracing but more general techniques such as a finite difference time-domain (FDTD) method that evolves Maxwell's equation can be foreseen in the future \cite{StijnARENA2018,Deaconu:2018bkf}. In the ray-tracing approximation, the different ray paths connecting an emitter and receiver can be classified as \emph{direct}, if the depth is monotonously decreasing or increasing along the path between emitter and receiver, as \emph{refracted}, if the path shows a turning point, and as \emph{reflected}, if the ray is reflected off the ice-air interface at the surface which acts as a perfect mirror for most geometries. A few typical ray-tracing solutions are presented in Fig.~\ref{fig:raytracing}.

\subsection{Analytic ray tracing}
\label{sec:analytic_rays}
The default signal propagation module in \NuRadioMC\ is an analytic ray-tracing technique that provides an unprecedented combination of speed and precision relative to traditional ray-tracing techniques. 
Traditional ray-tracing techniques locate the path connecting an emitter and receiver by time intensive trial-and-error methods, where numerous rays are ``thrown'' until a ray which connects the emitter and receiver is found. This is necessary because the index-of-refraction ($n$) of glacial ice is known to vary with depth, and so a light ray is bent and follows a curved path as it travels from an emitter to a receiver. Because the index-of-refraction does not need to be a well-behaved function it is impossible to predict the path traversed by the ray with full generality.

However, ice density measurements and the resulting index-of-refraction profiles from the South Pole and Moore's Bay site exhibit a simple, depth-dependent index-of-refraction $n(z)$. The data can be described to within a few percent \cite{Barwick_2018} by an exponential function of the following form:
\begin{equation}
n(z) = n_\mathrm{ice} - \Delta_n e^{z/z_0} \, ,
\label{eq:n_exp}
\end{equation}
where $z$ is the depth and $n_\mathrm{ice}, \Delta_n, z_0$ are the parameters of the model.
For this specific exponential $n(z)$ profile, an \textit{analytic} solution of the ray path as a function of depth ($y(z)$) exists and is given by
\begin{multline}
 y(z) = \pm z_0 \left(n^2_\mathrm{ice} C_0^2 -1\right)^{-1/2} \\ \cdot \ln\left(\gamma / \left[2  \sqrt{c(\gamma^2 - b \gamma + c)} - b\gamma + 2c \right]\right)+ C_1 \, ,
\label{eq:analytic_solution}
\end{multline}
with $\gamma = \Delta_n e^{z/z_0}$, $b = 2 n_\mathrm{ice}$, and $c = n_\mathrm{ice}^2 - C_0^{-2}$. We provide a derivation of this equation in \ref{sec:analytic_ray_tracing_derivation}.
The parameters $C_0$ and $C_1$ 
uniquely describe the ray path and need to be determined from two initial conditions which are given by the two points the ray goes through, e.g., the neutrino interaction vertex (the point of emission) and the observer position.  

The parameter $C_1$ corresponds to a horizontal translation in the coordinate system and can be calculated analytically from the initial conditions. The parameter $C_0$ must be determined numerically, and is found through a least-squares minimization.
For each receiver-emitter coordinate pair, we can either have no, one or two solutions, corresponding to no connecting ray, one connecting ray, or two connecting rays. To quickly and stably find all possible solutions, we leverage numerical algorithms as documented in \ref{sec:analytic_ray_tracing_fit}.

\subsubsection{Derived quantities}
Once a ray path is found, several derived quantities are needed in the simulation. The \emph{launch vector} of the ray is needed to calculate the \emph{viewing angle} (the angle between shower axis and launch vector) which is required to calculate the Askaryan emission. The \emph{receive vector} is needed to evaluate the antenna response for the arrival direction of the incident radiation. As discussed in \ref{sec:analytic_ray_tracing_fit}, the ray-tracing problem can be reduced to the y-z plane with a simple coordinate rotation. Hence, only the launch and receive angles are required, which can be calculated analytically from the derivative $\mathrm{d} y(z)/\mathrm{d} z$ which we specify in appendix \ref{sec:analytic_ray_tracing_derivative}.

The path length can be calculated numerically via the following line integral
\begin{equation}
 d = \int\limits_{z_1}^{z'_2} \left|\frac{d\vec{x}}{dz}\right| dz = \int\limits_{z_1}^{z'_2} \sqrt{1 + \left(\frac{d y(z)}{dz}\right)^2} \, dz \, ,
\end{equation}
where $\vec{x} = (y(z), z)^T$, and $z_1/2$ refer to the z position of the emitter/receiver. In case of a direct ray we have $z'_2 = z_2$. In case of a refracted or reflected ray, we first need to integrate from $z_1$ to the turning point and then the same path backwards to $z_2$. 

Similarly, the travel time $t$ and the signal attenuation $\exp(-A)$ can be calculated as
\begin{equation}
 t = \int\limits_{z_1}^{z'_2} n(z) / c \left|\frac{d\vec{x}}{dz}\right| dz = \int\limits_{z_1}^{z'_2} n(z) / c\sqrt{1 + \left(\frac{d y(z)}{dz}\right)^2} \, dz \, ,
\end{equation}
and
\begin{align}
 A & = \int\limits_{z_1}^{z'_2} \left|\frac{d\vec{x}}{dz}\right| / L(z, f) dz \\ & =  \int\limits_{z_1}^{z'_2} \sqrt{1 + \left(\frac{d y(z)}{dz}\right)^2} \, /L(z, f) \, dz \,
\end{align}
where $L(z, f)$ is the attenuation length as a function of depth and frequency which is discussed in Sec.~\ref{sec:attenuationlength}.

If the index of refraction profile is described with an exponential function as in Eq.~\ref{eq:n_exp}, an analytic expression for the path length and travel time can be derived. This analytic function is used by default due to its improved computing time. The derivation can be found in \ref{appendix:analyticT}. For the attenuation factor no analytic solution has been found and a numerical integration is required.

\subsubsection{Computational speed}
We provide a Python implementation of the analytic ray-tracing technique described above which leverages the NumPy \cite{derWalt:2011} and SciPy \cite{Jones:2001} computational packages. In addition, we implemented the time critical operations of finding the ray-tracing solution and determining the signal attenuation in a standalone C++ module. This C++ module leads to a substantial speed improvement of a factor of 20, so that the calculation of the ray-tracing solutions and the calculation of travel time and distance as well as the signal attenuation takes less than \SI{4}{ms} in ice. The C++ module utilizes the highly optimized and broadly supported GNU Scientific Library (GSL) \cite{GSL} for numerical integration and root-finding.

We provide a Cython wrapper to the C++ implementation so that it can be called as a sub-routine.  Selection of routine (C++ or Python) is done in a transparent fashion. If the user compiled the C++ extension, \NuRadioMC\ will automatically pick the faster C++ implementation, and otherwise utilize the Python implementation. In this way, the \NuRadioMC\ code works out-of-the-box without additional dependencies. The Python implementation is still sufficiently fast to be used for many problems.

\subsection{Focusing effect due to ray bending}
Applying the ray approximation to signals from neutrinos in case of ray bending, requires an additional correction factor on the signal amplitude. 
In general, when considering many rays which are bent there can either be a convergence or divergence of rays. If there is a convergence the ray density and thereby the amplitude of the signal will increase, and conversely so for a divergence. For the ice geometry, refraction contains the signal within the ice, and an amplification is expected if the receiver is above the point of emission and the ray is not reflected from the surface.

We calculate a correction factor from an energy conservation argument (please see \cite{Bouma:2025qvh} for a more detailed discussion): 
The intensity along the ray is given by 
\begin{equation}
I = \sqrt{\frac{\epsilon}{\mu}} \frac{\varepsilon^2}{c} = \frac{n \, \varepsilon^2}{c} \, ,
\end{equation}
for $\mu = \mu_0$ where $\varepsilon$ is the electric-field amplitude, $c$ the speed of light and $n$ the index of refraction. The total power contained in a ray bundle is $P = I \, A$ with $A$ being an area perpendicular to the propagation direction, so the electric field strength propagates as 
\begin{equation}
    \varepsilon' = \varepsilon \sqrt{\frac{n}{n'} \frac{A}{A'}} \, .
\end{equation}
The power radiated into a given solid angle is fixed by the source. In the case of  straight-line propagation:
\begin{equation}
\mathrm{d}A = R^2 \mathrm{d}\Omega = R \, \mathrm{d}\theta \times R \, \sin\theta\, \mathrm{d}\phi  \, .
\end{equation}
For refracted rays the relation $\frac{\mathrm{d}A}{\mathrm{d}\Omega}$ changes during propagation. Assuming a planar index of refraction model, i.e. $n(z)$ only depends on the depth $z$, the $R \, \mathrm{d}\theta$ part changes and is given by $\frac{\mathrm{d}z}{\mathrm{d}{\theta}} \sin\theta'$.
As $\mathrm{d} \phi$ remains constant along the ray bundle, the width in the $\phi$-direction 
is given by $r \, \mathrm{d} \phi$, where $r$ is the horizontal distance between the start and end points. The perpendicular area of the ray bundle at the receiver is therefore:
\begin{equation}
    \mathrm{d}A' = \frac{\mathrm{d}z}{\mathrm{d}{\theta}} \sin\theta' \, \mathrm{d}\theta \times r \, \mathrm{d} \phi.
\end{equation}
See \ref{appendix:focusing} for a derivation of this relation. Then, the ratio of electric field amplitudes is given by
\begin{equation}
    \frac{\varepsilon'^2}{\varepsilon^2} = \frac{n}{n'}  \frac{\mathrm{d}A}{\mathrm{d}A'} = \frac{n}{n'} \frac{R}{\sin\theta' \frac{\mathrm{d}z}{\mathrm{d}{\theta}}} \frac{R \sin\theta}{r}\,
\end{equation}
which is applied as a correction factor to the calculated electric field amplitude from the signal generation module. 
The factor $\frac{\mathrm{d}z}{\mathrm{d}{\theta}}$ is calculated numerically using the ray tracing code by calculating a new ray to the receiver position which is vertically displaced by a small amount $\Delta z \approx \SI{1}{cm}$.

Emitter positions very close to the shadow zone boundary require special attention as the correction diverges because $\frac{\mathrm{d}z}{\mathrm{d}{\theta}}$ approaches zero. This is not physical but an artifact from treating both emitter and receiver as a point. However, in reality the emission region is extended over several meters due to the extent of the particle shower (cf. Fig.~\ref{fig:ARZ_EM16}) and also the antenna is an extended object. Thus, we studied the stability of the correction factor under small changes of the emitter position by $\pm\SI{5}{m}$ corresponding to typical dimensions of the emission region. We find that correction factors below about a factor of 2x in amplitude vary by less than 10\% when the emitter position is varied. Larger amplification factors in-turn are not stable. Hence, limiting the amplification to a maximum of 2x removes unphysical correction factors.
Furthermore, we studied the effect of the limit value. Limiting the focusing correction to a factor of 1.5x, 2x and 3x results in essentially the same effective volume (i.e. sensitivity of the detector) over a broad range of neutrino energies. Thus, the exact choice limit value is not that important as long as very large amplification factors are removed. As default we limit the focusing correction to a factor of 2x but allow the user to configure this value via the config file.

The effect of focusing is strongest when the rays pass near the surface and experience significant refraction. For a receiver close to the surface we find an increase in the effective volume of the order of 10\% due to this correction. 

\subsection{Numerical ray tracing for arbitrary density fields}
In the future, it may become necessary to describe the ice in more detail than an exponential profile that only depends on the depth. This will require a more detailed ray tracing that takes into account an arbitrary 3D index of refraction profile $n(x, y, z)$. We have already foreseen this case and ensured that necessary hooks are available in the code.

Interestingly, the computational problem of the propagation of ultra-high energy cosmic rays through the universe is similar to propagating a ray through the ice. Instead of magnetic fields bending the trajectories of charged cosmic-ray particles, the ray is bend according to the spatial distribution of
the index of refraction.  Where the cosmic ray can spallate into secondaries, a ray can be partly transmitted and reflected. Consequently, we considered the cosmic-ray propagation code CRPropa \cite{CRPropa} as one option and have started to modify it for our needs.

The resulting code RadioPropa~\cite{RadioPropa} solves the Eikonal equation in a local paraxial approximation thus enabling casting of rays through materials with arbitrary varying refractive index as may be required here.  In addition, RadioPropa handles effects from boundary traversals such as reflection or
partial reflection and allows for the implementation of propagating components of the electric field differently, such as needed for birefringence.  It automatically tracks several parts of the original ray, making it also suitable for other less well understood phenomena in the ice. In the same way as \NuRadioMC, RadioPropa is modular and flexible, leaving room for future developments. It is currently under heavy development and therefore not yet fully included in \NuRadioMC.

\subsection{Signal propagation beyond ray tracing}

Ray tracing describes the path taken by light in the limit where
the wavelength is much smaller than any relevant feature sizes. While this is
appropriate in most practical cases, i.e., when the ice is uniform or has a slowly-varying index of refraction, ray tracing does not offer a full description of light propagation near dielectric interfaces, where additional solutions to Maxwell's equations exist, (see e.g.~\cite{Frezza:15} for a pedagogical tutorial on some of the solutions, or \cite{planar_track} for a complete solution for the field of a particle track).  In addition to the ice-air interface at the surface, variations in ice density are present below the surface, producing a set of dielectric interfaces. These may result in signals being observed at locations, where simple models assuming a smooth gradient predict no radio signals \cite{Barwick_2018}. While adaptations to the analytic ray-tracing requiring a smooth gradient of the index of refraction, deliver solutions for special cases, the finite-difference time-domain (FDTD) method may be used to model propagation in ice even in the presence of inhomogenities in all its aspects \cite{Deaconu:2018bkf,StijnARENA2018}. 

Interesting phenomena that arise include the existence of potentially detectable (though generally small) signals coming from regions where there is no ray-tracing solution, diffraction and interference of the radio waves, and the presence of caustics, where the small electric field may be significantly amplified in some geometries \cite{Deaconu:2018bkf}.

While these effects will slightly modify the effective volume of a detector
and provide additional opportunities for event reconstruction, direct integration
of an FDTD solver into \NuRadioMC\ is challenging for the purpose of providing a simulation framework. 
FDTD methods are very computationally and memory intensive, requiring
discretization on the scale of a tenth of the smallest relevant wavelength in
all spatial dimensions as well as time. Directly simulating the entire volume
seen by a typical in-ice station is extremely computationally challenging in
three dimensions with our present capabilities -- we estimate a single simulation
of a cubic kilometer volume valid up to 500 MHz would take $\mathcal{O}(10^7)$
CPU-hours. One can envision the usage for a single event (in case of re-simulation of a detected shower for example), the integration for all events is, however, impractical. 

By considering only azimuthally-symmetric antennas and density variations
dependent only on depth, it is possible to simulate a transmitting in-ice
antenna in just two dimensions, greatly reducing the computational burden. We
are investigating techniques exploiting reciprocity in order to tabulate the
propagation properties of the equivalent time-reversed geometry, corresponding
to a receiving antenna. Such tabulated properties could then be incorporated
into \NuRadioMC\ in an efficient manner. 


\section{Detector simulation}

The fourth pillar of \NuRadioMC\ is the detector simulation, i.e., the calculation of the detector response to an electric field at the antenna and subsequent trigger simulation. We use the software \emph{NuRadioReco} for this task \cite{NuRadioReco}. NuRadioReco is a software for the detector simulation and event reconstruction of radio neutrino and cosmic-ray detectors. It is written in Python and also follows a modern modular design so that it nicely integrates into \NuRadioMC. 

\subsection{Antenna simulation}

The most important part in the simulation of the detector response is the impact of the antenna. NuRadioReco provides antenna response pattern of typically used antennas such as LPDAs, dipoles or bicone antennas that were simulated with dedicated codes such as WIPL-D \cite{Kolundzija2011} and XFDTD \cite{XFDTD}. NuRadioReco also provides an interface to the output of these codes such that new antenna models can be added if necessary. 

In earlier software, the response of the antennas was typically treated in a simplified way, only assuming real gain factors and a simple polarization response, i.e.~ignoring contributions polarized orthogonal to the main antenna sensitivity. According to methods already standard in the treatment of radio signal from cosmic rays (e.g. \cite{RadioOffline,LOFARCr}), the antenna response is modelled fully frequency-dependent in NuRadioReco, also taking into account the group delay induced by the antenna and its sensitivity to two orthogonal polarization components. 

\subsection{Trigger simulation}

Especially when looking for small signals, as expected from neutrinos, the simulation of the trigger mechanism is essential. The trigger simulation is set up as such that any instrumental trigger can be rebuilt in software. NuRadioReco offers modules to simulate different trigger conditions, e.g., a simple threshold trigger, a high and low trigger as implemented on the SST electronic \cite{Kleinfelder:2015ipa} used by ARIANNA \cite{ARIANNA2015} that also allows to specify temporal coincidences between different channels, or  more complex triggers such as the phased array concept used by ARA \cite{ARAprogress} have been included to model the instrument response as implemented in the fields. 

\subsection{Usage in complex detectors}
\label{sec:detectordescription}

NuRadioReco was built to reconstruct data from an existing detector. In order to facilitate complex detectors without creating too much overhead, the detector description is stored in a database allowing for a description of every single detector component. While this functionality will be helpful to simulate specific events for an existing detector, it is much too complex for design studies. Therefore, NuRadioReco also allows the user to define the detector description in a human readable JSON format, with reduced complexity. This means both that the detector description only needs to be as complex as minimally required and it significantly speeds up simulations. The information ranges from basic parameters such as the positions of the antennas, their type and orientation to more detailed properties such as the sampling rate of the digitizing electronics, the cable lengths or details about the amplifier and ADC. The detector simulation modules have access to these properties and will simulate the detector response accordingly. An example of a typical detector simulation is provided in \ref{sec:detector_sim}.

\section{Utilities}
The four pillars of \NuRadioMC\ are complemented by a set of utility classes that are available to all modules, such as units and medium properties. 

\subsection{Cross-sections and inelasticities}
The cross-section of neutrinos at energies relevant for radio detection are still subject to study, given that these energies have never been probed. Different current extrapolations \cite{Gandhi1998,Connolly2011,CooperSarkar:2011pa} have been implemented in \NuRadioMC\ in the central utilities, so that the cross-sections can easily be exchanged throughout the code, if so desired.

\subsection{Earth models for neutrino absorption}
\label{sec:earthattenuation}
To simulate the sensitivity of a neutrino detector, we need to calculate the probability of a neutrino reaching the detection volume. The Earth atmosphere has negligible absorption for high energy neutrinos but the Earth becomes opaque at high neutrino energies. Hence, \NuRadioMC\ comes with multiple models to calculate the Earth absorption so that we can assign each simulated neutrino a weight, i.e., a probability of reaching the detection volume.

Right now, \NuRadioMC\ provides two Earth models: a simple Earth model with a constant density and a core-mantle-crust Earth model with three layers of different densities. Due to the modularity, it is straight forward to add more detailed models if deemed necessary. 

Currently, we do not model \emph{tau regeneration}: A tau lepton that is created following a tau neutrino interaction can propagate significantly through the Earth and potentially decay with a relatively large energy and producing another tau neutrino that can interact close to the detector. We plan to include this effect in a future version of \NuRadioMC\ using e.g. the code of \cite{Alvarez-Muniz2018,Alvarez-Muniz2019}.

\subsubsection{Simple Earth model}
This model uses a constant density of \SI{2900}{kg/m^3} and by default uses the cross section ($\sigma$) based on \cite{Gandhi1998}. It then calculates the distance the neutrino goes through the Earth as
\begin{equation}
d = 2R_{e}cos(\pi - \vartheta),
\end{equation}
where $R_e$ is the radius of the Earth and $\vartheta$ is the zenith angle of the neutrino direction. The weight of an event is then
\begin{equation}
\mathrm{weight} = e^{-d \sigma \rho / AMU},
\end{equation}
where $\rho$ is the constant density of the Earth and AMU is the atomic mass unit in kg.

\subsubsection{Core-mantle-crust Earth model}
\NuRadioMC\ provides a more realistic Earth model with three layers of different densities which is the default model. In this model, the cross section is per default calculated based on \cite{Connolly2011} and the propagation distance is calculated through three different layers. The weight is calculated as
\begin{equation}
\mathrm{weight} = e^{-(d_{1}\rho_{1} + d_{2}\rho_{2} + d_{3}\rho_{3}) \sigma / AMU},
\end{equation}
where \(d_{1}\), \(d_{2}\), \(d_{3}\) are the distances through three layers and \(\rho_{1}\), \(\rho_{2}\), \(\rho_{3}\) are the three densities.

\subsection{Handling of Fourier Transforms}
\NuRadioMC\ provides a consistent internal handling of Fourier transforms. A common source of errors when using time- and frequency-domain calculations simultaneously is the normalization of the Fourier transforms. There are several reasons for different normalizations depending on the purpose and context. All \NuRadioMC\ Fourier transforms adhere to Parseval's theorem and previously existing Askaryan signal parameterizations have been adjusted to match the FFT definition used in \NuRadioMC. Details are discussed in \ref{sec:FFT}. 

\subsection{Handling of units}
In simulations, typical errors occur during the handling of units. To prevent that, \NuRadioMC\ (just like NuRadioReco) employs a default system of units, a concept borrowed from the Pierre Auger Observatory offline analysis framework~\cite{OfflineSoftware}: every time a physical variable is defined, it is multiplied by its unit, and every time a variable is plotted or printed out in a certain unit, it is divided by the unit of choice. All other calculations within the code can then be done without considering units. 
\begin{minted}[
fontsize=\footnotesize
]{python}
from NuRadioMC.utilities import units
time = 132. * units.ms # define 132 milliseconds
distance = 5. * units.mm # define 5 mm
speed = distance/time  
print("the speed is {:.2f} km/h"
    .format(speed/units.km*units.hour))
# the speed is 0.14 km/h
\end{minted}
The units utilities are available to modules written in both Python and C. In order to facilitate this, no standard Python package was used.

\subsection{Attenuation length and other medium characteristics}
\label{sec:attenuationlength}
As discussed in Sec.~\ref{sec:propogation} the signal propagation is a significant part of the neutrino simulation and an area where lots of development is still to be expected. Consequently characteristics of the interaction medium are stored centrally in the utilities to avoid contradicting definitions in modules. We describe the index-of-re\-frac\-tion profile and signal attenuation properties separately to allow for simulation with different combinations of the two. Which model is being used in a \NuRadioMC simulation is controlled via the central config file (see Sec.~\ref{sec:config}). 

Currently, a signal attenuation model for South Pole ice is provided that is based on a custom model used by the ARA experiment \cite{ARA_atten}. For the index-of-refraction profile we provide exponential parameterizations to data from for the South Pole and Moore's Bay \cite{Barwick_2018}, as well as from Greenland \cite{2008JGlac..54..839H,alley_koci_1988}.

\subsection{Flux calculations and sensitivity limits}
In order to compare the performance of different experimental designs, typically quantities like the effective area, volume or expected limits are compared. Since also here, many definitions are common (e.g. 90\% confidence upper limits vs.\ $5\sigma$ discovery fluxes), utility functions are provided centrally.

\section{Example 1: Calculation of the sensitivity of an Askaryan neutrino detector}
\begin{figure}[t]
 \centering
 \includegraphics[width=0.48\textwidth]{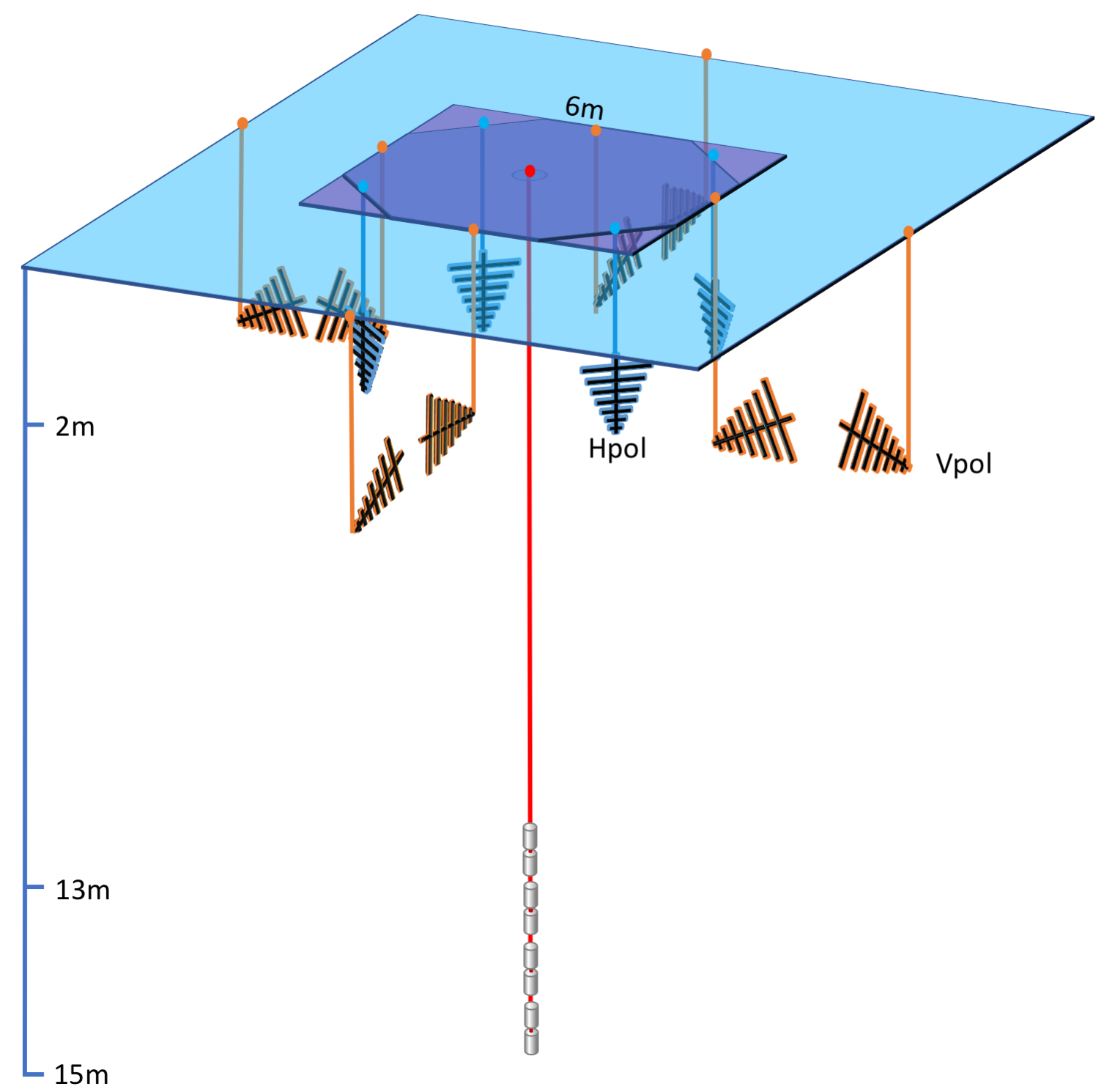}
 \caption{Sketch of the station layout simulated in Example 1.}
 \label{fig:station_layout}
 \end{figure}

In this section we present a full example of the capabilities of \NuRadioMC\ to simulate the sensitivity of an Askaryan detector. We choose a station layout that combines log-periodic dipole antennas (LPDA) near the surface with slim dipoles deployed in a borehole deeper in the ice. The specific layout is depicted in Fig.~\ref{fig:station_layout}. This station layout does not necessarily reflect the authors' opinion on the optimal detector layout but was chosen because it highlights \NuRadioMC's capabilities: Antennas of different type, orientation and depth are simulated, the location close to the surface makes a detailed propagation of the signal through the firn necessary, and multiple trigger conditions need to be calculated for different sets of antennas. In the following, only the relevant code snippets are shown. A comprehensive tutorial can be found online \cite{NuRadioMC_tutorial}.

\subsection{Event generation}
The first step in the simulation is the event generation. The event generation is done stand-alone and produces a list of neutrino interactions in the ice with all necessary properties saved in a simple HDF5 format (see Sec.~\ref{sec:event_generation} for details and advantages of separating this step). We choose to generate several input lists, each for a fixed neutrino energy to study the energy dependence. We only consider the initial neutrino interaction. A discussion of the impact of additional Askaryan signals from decaying taus or interacting muons goes beyond the scope of this publication. 

A list of one million neutrino interactions with an energy of $E_\nu = \SI{e18}{eV}$ in a cylindrical volume saved in chunks of 10,000 events can be generated with

\begin{minted}[
%frame=lines,
%framesep=2mm,
%baselinestretch=1.2,
fontsize=\footnotesize,
%linenos
]{python}
generate_eventlist_cylinder('1e18_n1e6.hdf5',
    n_events=1e6, n_events_per_file=1e4,
    Emin=1e18 * units.eV, Emax=1e18 * units.eV,
    fiducial_rmin=0, fiducial_rmax=5 * units.km,
    fiducial_zmin=-2.7 * units.km, fiducial_zmax=0)
\end{minted}
The radius needs to be set large enough to include all events that can trigger the detector and is set to \SI{4}{km} here. For larger neutrino energies, the radius needs to be extended and for lower energies the simulation volume can be decreased to save computing time. The vertical extent of the volume ranges from the surface to the bottom of the ice layer at a depth of \SI{2.7}{km} at the South Pole.

\subsection{Configuration of simulation parameters}
\label{sec:config}
The settings of the simulation are controlled with a config file in the human-readable \texttt{yaml} format. The user only needs to specify a parameter if it should be different from its default value. An example configuration with typical settings is shown in listing \ref{lst:config}. Typical parameters are the choice of signal generation model (\emph{Alvarez2009} in this example), the ice model, or if noise should be generated and added to the signal in the simulation. 

\begin{listing}[ht]
\begin{minted}[
%frame=lines,
%framesep=2mm,
%baselinestretch=1.2,
fontsize=\footnotesize,
%linenos
]{yaml}
noise: False  # specify if simulation should be 
    run with or without noise
sampling_rate: 5.  # sampling rate in GHz used 
    #internally in the simulation.
speedup:
  minimum_weight_cut: 1.e-5
  delta_C_cut: 0.698  # 40 degree
propagation:
  ice_model: southpole_2015
signal:
  model: Alvarez2009
trigger:
  noise_temperature: 300  # in Kelvin
weights:
  weight_mode: core_mantle_crust # core_mantle_crust: 
    #use the three layer earth model, 
    #which considers the different densities of the 
    #core, mantle and crust. 
    #Simple: use the simple earth model, 
    #which applies a constant earth density

\end{minted}
\caption{Example of \NuRadioMC's config file. All parameters are specified in a default system of units. See text for details.}
\label{lst:config}
\end{listing}

\subsection{Detector description}
The detector description consists of two parts. First, we need to define the layout of the detector (position, type, and orientation of the antennas), and the sampling rate. Additional parameters such as cable delays and amplifiers can be specified if needed (cf.~Sec.~\ref{sec:detectordescription} and NuRadioReco \cite{NuRadioReco}). However, in this example we will perform a simplified detector simulation sufficient to estimate the sensitivity of an Askaryan detector. The detector description is specified in a JSON file presented in List.~\ref{lst:detector}.

\begin{listing}[ht]
\begin{minted}[
%frame=lines,
%framesep=2mm,
%baselinestretch=1.2,
fontsize=\footnotesize,
%linenos
]{json}
{
    "channels": {
        "1": {
            "station_id": 101,
            "channel_id": 0,
            "ant_type": "createLPDA_100MHz",
            "ant_position_x": 3,
            "ant_position_y": 0,
            "ant_position_z": -2.0,
            "ant_rotation_phi": 180,
            "ant_rotation_theta": 90,
            "ant_orientation_phi": 0,
            "ant_orientation_theta": 180,
        },
        ...
    },
    "stations": {
        "1": {
            "pos_altitude": 0,
            "pos_easting": 0,
            "pos_northing": 0,
            "pos_site": "southpole",
            "station_id": 101
        }
    }
}
\end{minted}
\caption{Example of detector description. Only the first channel is shown which defines a downward facing LPDA at a depth of \SI{2}{m} with its tines oriented along the Northing direction.}
\label{lst:detector}
\end{listing}

Second, we need to specify basic details of the signal chain, i.e., what filter is being used and which triggers are calculated. These tasks are done by dedicated NuRadioReco modules \cite{NuRadioReco} (see Sec.~\ref{sec:detectordescription}) that interface directly with \NuRadioMC. Instead of simulating just a single trigger condition as shown in the example, a separate trigger can be simulated for each parallel pair of LPDA antennas and for the dipole antennas. This is achieved by calling the same trigger module several times with different arguments. The full example can be found in the online tutorial \cite{NuRadioMC_tutorial}. 

\subsection{Running the simulation, results, and visualization tools}
The \NuRadioMC\ simulation is run by executing the steering script from the command line. The flexibility to split up the input data set into smaller chunks is part of the event generator, so multi-processing computing resources can be used right away. A detailed example on how to run \NuRadioMC\ on a cluster is available in the online tutorial \cite{NuRadioMC-docu-cluster}. 

\begin{figure*}
    \centering
     \includegraphics[width=0.45\textwidth]{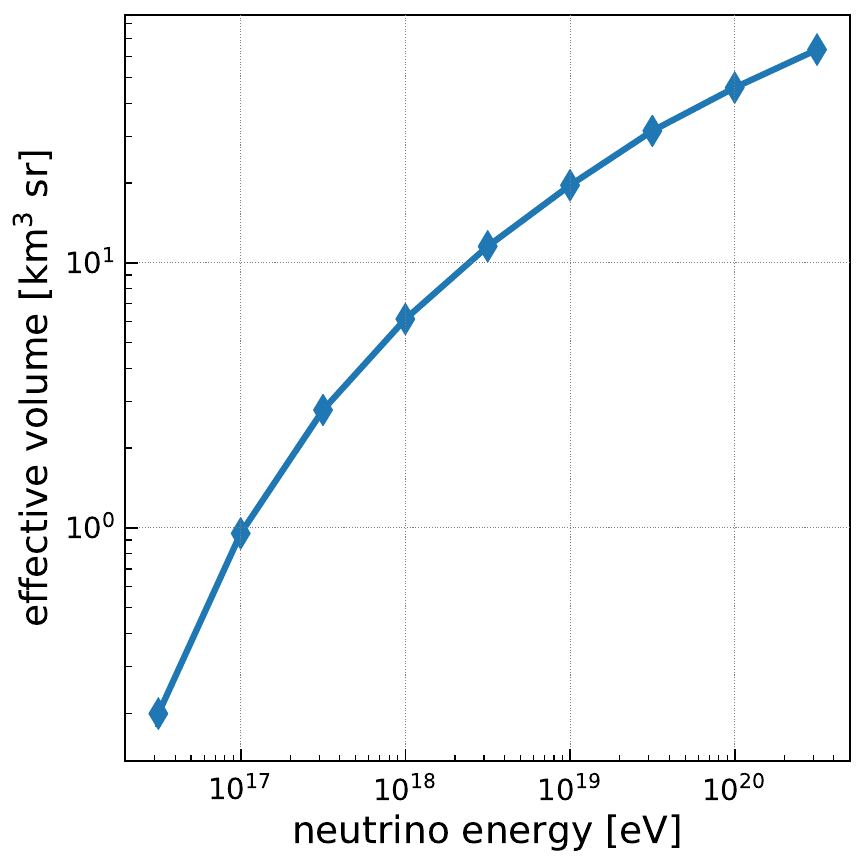}
    \includegraphics[width=0.54\textwidth]{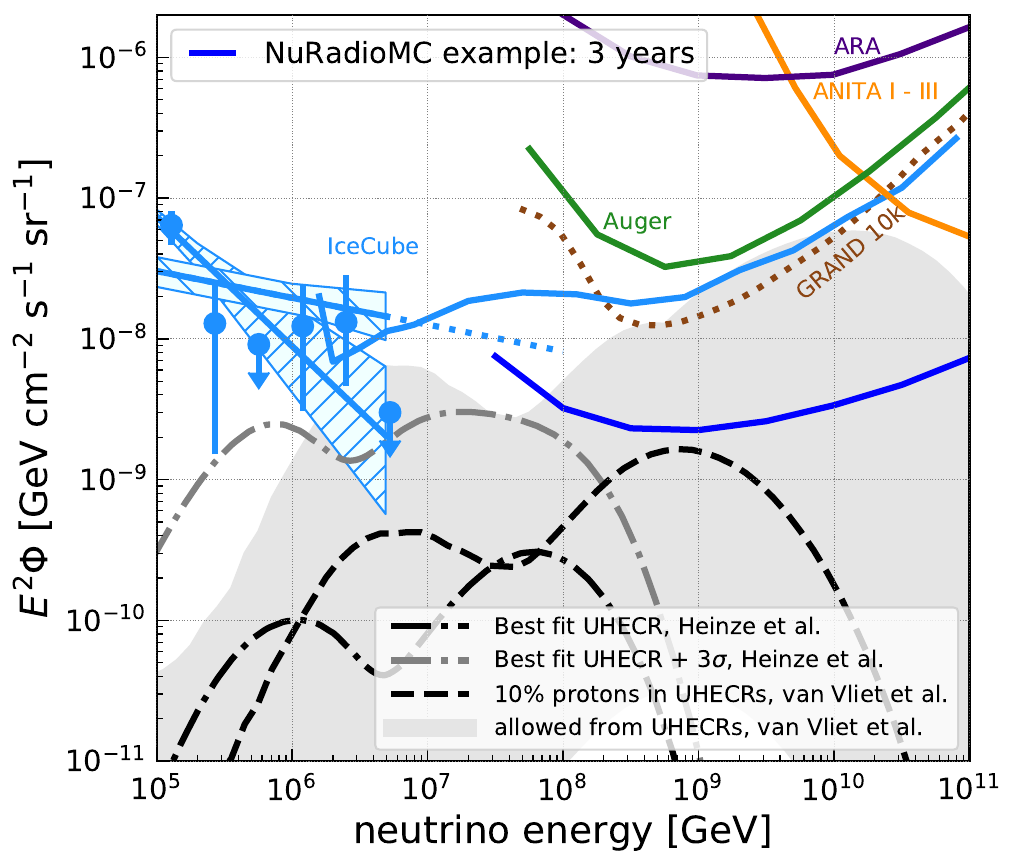}
    \caption{(left) effective volume of one example detector station (right) corresponding expected limit for a diffuse neutrino flux for a detector comprising 100 stations and an uptime of 3 years. Shown are for comparison neutrino flux measurements from IceCube \cite{Haack:2017dxi,Kopper:2017zzm,Aartsen:2018vtx}, the Pierre Auger Observatory \cite{Aab:2015kma}, ANITA \cite{Allison:2018cxu}, and ARA \cite{ARA}, as well as neutrino flux prediction models from \cite{Heinze:2019jou,vanVliet:2019nse} calculated using the restrictions from ultra-high energy cosmic rays. We also compare to other proposed arrays \cite{GRAND}.}
    \label{fig:Veff}
\end{figure*}

The sensitivity of the detector is quantified in terms of effective volume to an isotropic neutrino flux. It is given by the weighted sum of all triggered events divided by the total number of events multiplied by the simulation volume and the simulated solid angle (typically $4\pi$). The weighting factor is the probability of a neutrino reaching the simulation volume (and not being absorbed by the Earth). The effective volume of our example detector station is presented in Fig.~\ref{fig:Veff} (left). This effective volume can be converted into an expected limit on the diffuse neutrino flux which is shown in the right panel of Fig.~\ref{fig:Veff}. The required tools to make these standard post-processing plots are also part of \NuRadioMC. 

\begin{figure}
    \centering
    \includegraphics[width=0.49\textwidth]{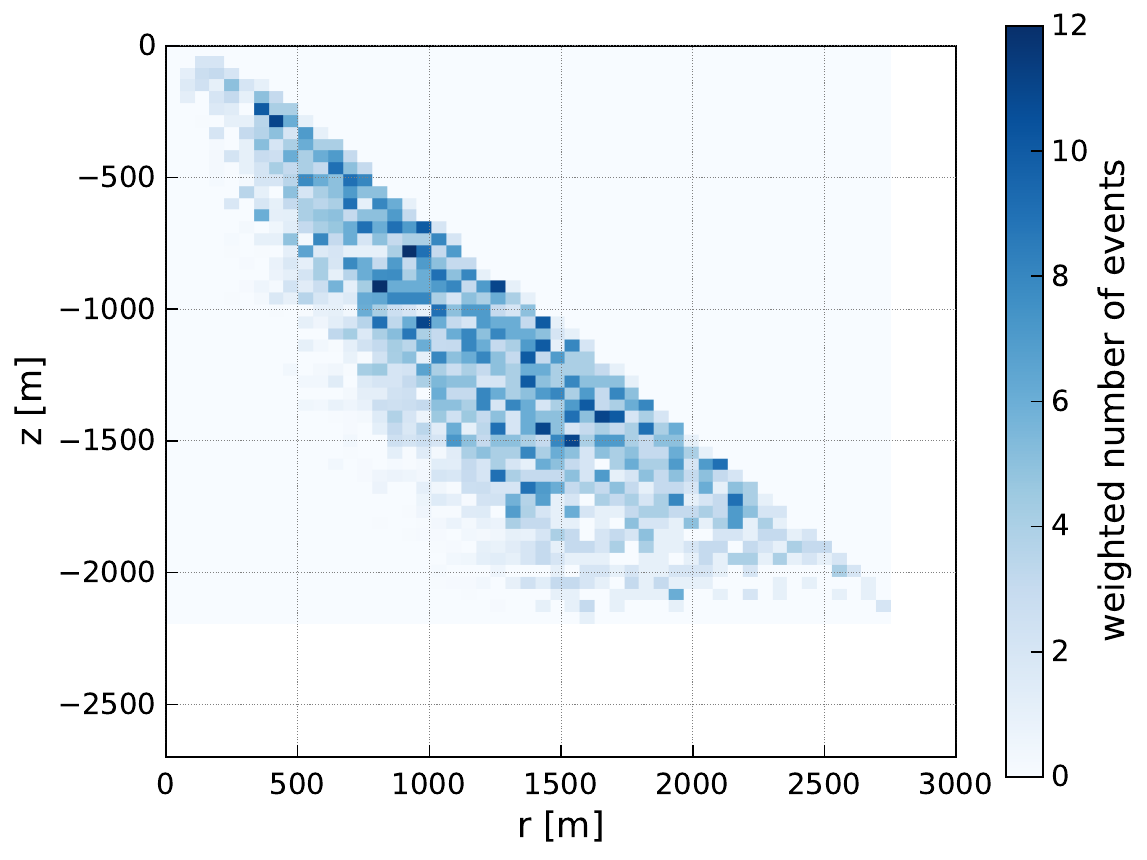}
    \includegraphics[width=0.49\textwidth]{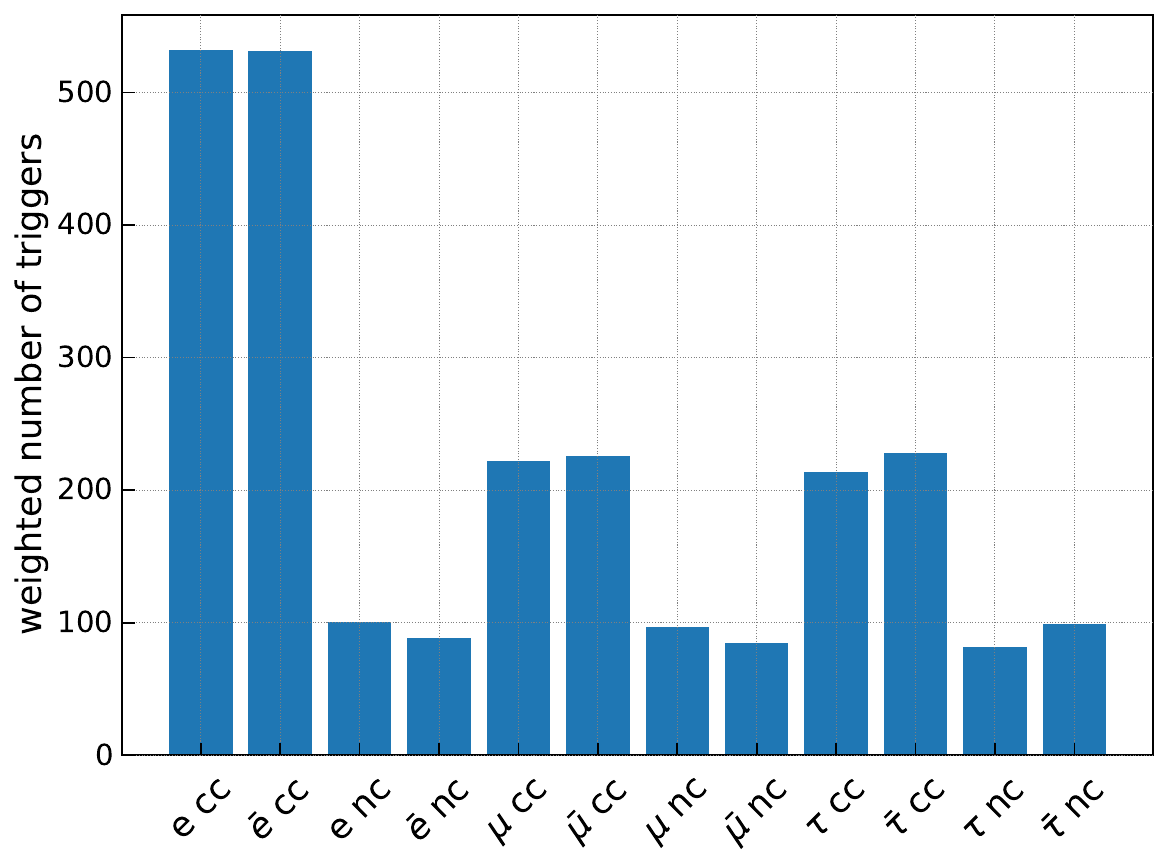}
    \caption{Visualization for the energy bin of \SI{e18}{eV} neutrino energy. (top) Distribution of neutrino interaction vertices of all triggered events. (bottom) Flavor and interaction type (charged or neutral current) distribution of triggered events.}
    \label{fig:vertex}
\end{figure}

Furthermore, a standard set of debug plots can be automatically generated from the output files. The distribution of the neutrino interaction vertices of events that triggered the detector is shown in Fig.~\ref{fig:vertex} (left). The upper right (triangular) part of the volume correspond to positions in the \emph{shadow zone} where signals cannot reach the detector according to the ray tracing. The lower left region has little events because the Askaryan signal is only emitted towards the antennas if the neutrino is up-going, i.e., it travelled through the Earth and its probability of reaching the detector is small. The right panel shows the ratio of neutrino flavors and interaction types that triggered the detector. In this case, most triggered events were electron neutrino charged-current (CC) interactions where the full neutrino energy is deposited in particle showers producing an Askaryan signal.

\section{Example 2: Calculation of the efficiency to detect a signal from both the direct and reflected path}
In this example, we calculate the efficiency of an in-ice antenna to observe both the direct Askaryan signal and the signal reflected at the ice surface. For most shower geometries there is total internal reflection of the Askaryan signal at the ice surface, i.e., the ice-air interface acts as a mirror. Consequently, an antenna installed within the ice has the chance to see two pulses: one pulse that propagated straight to the antenna and a second pulse that was reflected off the surface. Detecting this \emph{D'n'R} (direct and reflected) signature is advantageous and an Askaryan neutrino detector will benefit strongly from detecting both pulses: First, it provides a unique method to identify a neutrino interaction in the ice as origin of the detected radio signal, and second, the time difference between the two pulses allows for an improvement in the reconstruction of the distance to the neutrino interaction vertex which is a crucial ingredient for the reconstruction of the neutrino energy. See \cite{Allison2019} and \cite{ARIA} for first experimental results concerning this effect using pulsers deployed in the Antarctic ice at South Pole.

There are several effects that influence the efficiency of detecting both pulses that are all taken into account in the \NuRadioMC\ simulation: 
\begin{itemize}
    
\item The reflection coefficient depends on the incident angle of the radio pulse at the ice surface and can range from 1 (total internal reflection) to 0 (no reflection) at the Brewster angle. 
\item The reflection results in a phase shift of the Askaryan pulse which can alter the amplitude of the pulse. This is modelled using the complex Fresnel coefficients.
\item Due to the changing index of refraction in the upper ice layers the signal propagates on curved paths. We find all possible paths to each antenna via ray-tracing. We note that not only a 'direct' and 'reflected' path will provide a useful signature but any two distinct paths through the ice to the antenna. In case only one solution exists, the efficiency to detect two pulses is of course zero. 
\item The different ray paths correspond to different launch angles of the signal. This results in a potentially large difference of the amplitude of the Askaryan signal as the launch angles correspond to different viewing angles.
\item Antennas have a different sensitivity to different incoming signal directions. 
\item The two ray paths have different propagation distances and potentially propagate through ice with different attenuation lengths. 
\end{itemize}

In the following we describe an example of how to simulate the D'n'R detection efficiency with \NuRadioMC\ and explain the relevant parts of the code. The full code of this example can be found online at \cite{Example2}.

The D'n'R efficiency depends on the depth of an antenna, hence, we want to define a detector with several antennas of the same kind at different depths. As antenna type we choose a bicone antenna as used by the ARA experiment as such an antenna is sensitive to the dominant vertical polarization, fits into narrow boreholes, and has very little signal dispersion which helps to measure the time difference between the two pules. Hence, we set up a detector with vertically oriented bicone antennas every \SI{10}{m} down to a depth of \SI{100}{m}.

It does make sense to study the D'n'R efficiency as a function of neutrino energy. Therefore, we can use the same script to generate the input event list as in the previous example. 

\subsection{Set-up of detector simulation}

In the previous example we have discussed how to simulate the detector response and the trigger. In the detector simulation so far, all signals that reach the antenna from the different ray path solutions, are combined into a single voltage trace on which the trigger condition is determined. However, for the D'n'R study, we not only need to determine if the detector could observer/trigger a certain event, but also if both pulses are visible. Hence, a dedicated \emph{NuRadioReco} module called \emph{calculateAmplitudePerRaySolution} was written, which simulates the antenna response to each pulse separately and calculates and saves the resulting maximum amplitude. Following this we can calculate if a triggered events has two visible pulses.

As trigger condition we choose a simple threshold trigger of $2~V_\mathrm{rms}$ that runs on all channels (i.e.~antennas) independently. The \NuRadioMC\ simulation is then executed as described in Example 1.

\subsection{Results}
We now assume a more stringent cut in which all events that produce at least a $3\sigma$ ($3~V_\mathrm{rms}$) signal can be recorded. For the seconds pulse the requirement for identification is assumed smaller at $2\sigma$. Furthermore, we require that the time difference between the two pulses is smaller than 430 ns which is assumed as typical record length. We then calculate if an event has triggered via
\begin{equation}
B_i = A_1^i \geq 3~V_\mathrm{rms}\text{ or }A_2^i \geq 3~V_\mathrm{rms}
\end{equation}
and if both pulses are visible via
\begin{align}
C_i  = &((A^i_1 >= 3~V_\mathrm{rms}) \text{ or } (A^i_2 >= 3~V_\mathrm{rms}))\\
&\text{ and }((A^i_1 >= 2~V_\mathrm{rms})\text{ and }(A^i_2 >= 2~V_\mathrm{rms}))\\
&\text{ and }(\Delta T < \SI{430}{ns}) \, ,
\end{align}
where $A^i_1$ and $A^i_2$ are the amplitudes of the two pulses of event $i$.

Then the D'n'R efficiency is then given by
\begin{equation}
\epsilon = \sum_i C_i/ \sum_i B_i
\end{equation}
where the summation runs over all simulated events $i$. This calculation is performed for each simulated antenna depths, and for each set of simulated neutrino energy separately.

\begin{figure}[t]
    \centering
    \includegraphics[width=0.49\textwidth]{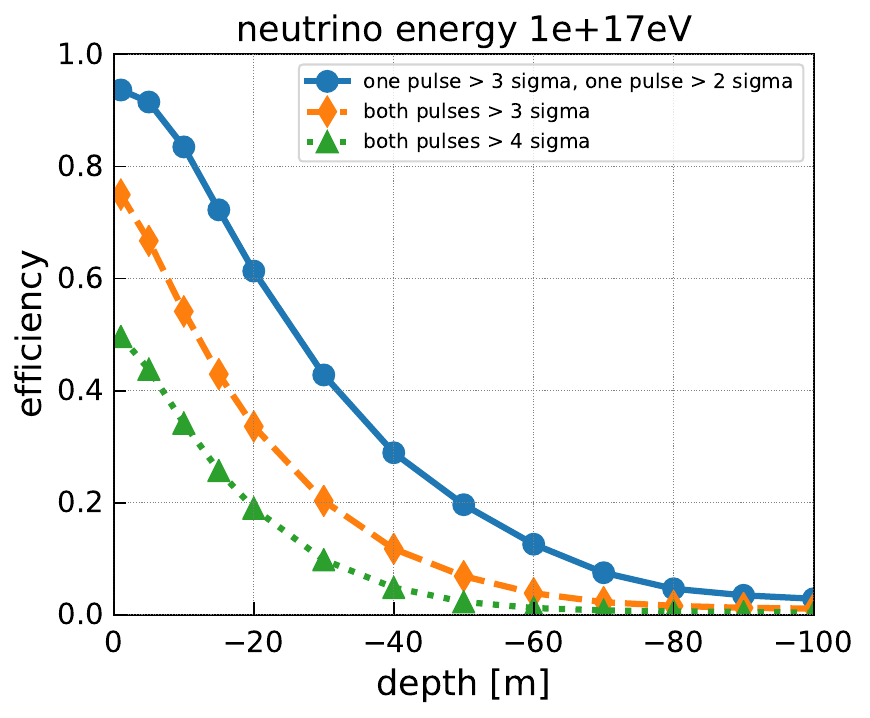}
    \includegraphics[width=0.49\textwidth]{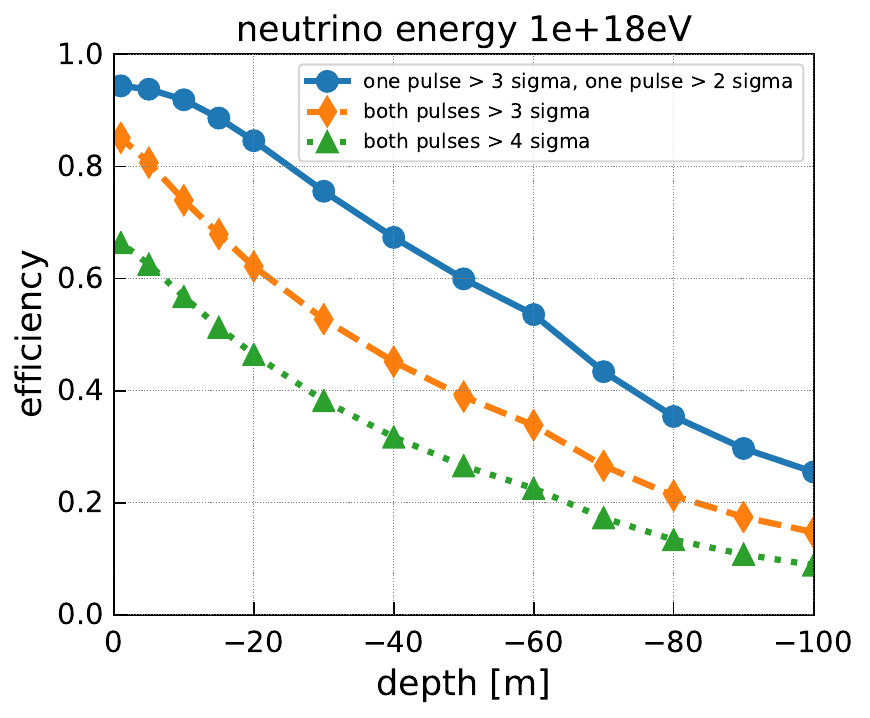}
    \caption{Efficiency to detect both the direct and reflected Askaryan pulse as a function of depth of the receiver. (top) For a neutrino energy of \SI{e17}{eV}. (bottom) For a neutrino energy of \SI{e18}{eV}. Different markers and colors correspond to different trigger thresholds. All events with a signal of at least a $3 \sigma$ in any of the pulses were considered which explains the smaller efficiency at the surface for the 'both pulses $> 4 \sigma$' criterion.}
    \label{fig:DnR}
\end{figure}

We simulated 10 million events per neutrino energy and obtain the result presented in Fig.~\ref{fig:DnR}. The D'n'R efficiency depends strongly on depth and energy and is best at shallow depth and high energies. 

It should be noted that D'n'R efficiency is not the only parameter that one should optimize an array for. For example, a shallower station generally has a smaller effective volume than a deep station, and the fraction of sky coverage also depends of depth. Together with a diverse choice of antennas influencing reconstruction capabilities, data volume restrictions, and instrument costing, optimizing a detector layout is a complex problem, for which \NuRadioMC\ provides guidance. 

\section{Example 3: Optimization of station spacing for an Askaryan neutrino detector}

In this example we calculate the probability to detect a signal from the same neutrino in multiple stations of an array. For a discovery detector, one objective is a large sensitivity which means that it is beneficial to separate stations far enough to minimize station coincidences. However, one may want to optimize differently in the future to have a large fraction of coincidences to improve reconstruction quality. Here, we show how the coincidence fraction can be studied as a function of station separation distance, neutrino energy, and antenna depth. The full code of this example can be found online at \cite{Example3}.

\subsection{Simulation strategy}

We consider a simplified detector with two components. The first one is a surface oriented component consisting of LPDAs and dipoles. To save computing time, we only simulate two orthogonally-oriented horizontal LPDAs at \SI{2}{m} depth and one dipole at \SI{5}{m} depth to be sensitive to all signal polarizations. The second component is a deep one, approximated with a single dipole antenna at \SI{50}{m} depth. We combine the four antennas into a single station so that only one simulation needs to be run, but we can still evaluate the coincidence fraction independently.  

In principle, one would need to simulate a full 2D grid for every station separation distance that one wanted to test, because there might be cases where not the nearest station triggered but the next-to nearest neighboring station or stations even further out. However, as this will drastically increase computing time (which scales linearly with the number of stations) this small second order effect is ignored in this example. Our analysis will show that the coincidence rate is dominated by the nearest neighbors, i.e., the coincidence rate quickly drops if the separation between stations is doubled, justifying this approximation.

\begin{figure}[t]
    \centering
    \includegraphics[width=0.45\textwidth]{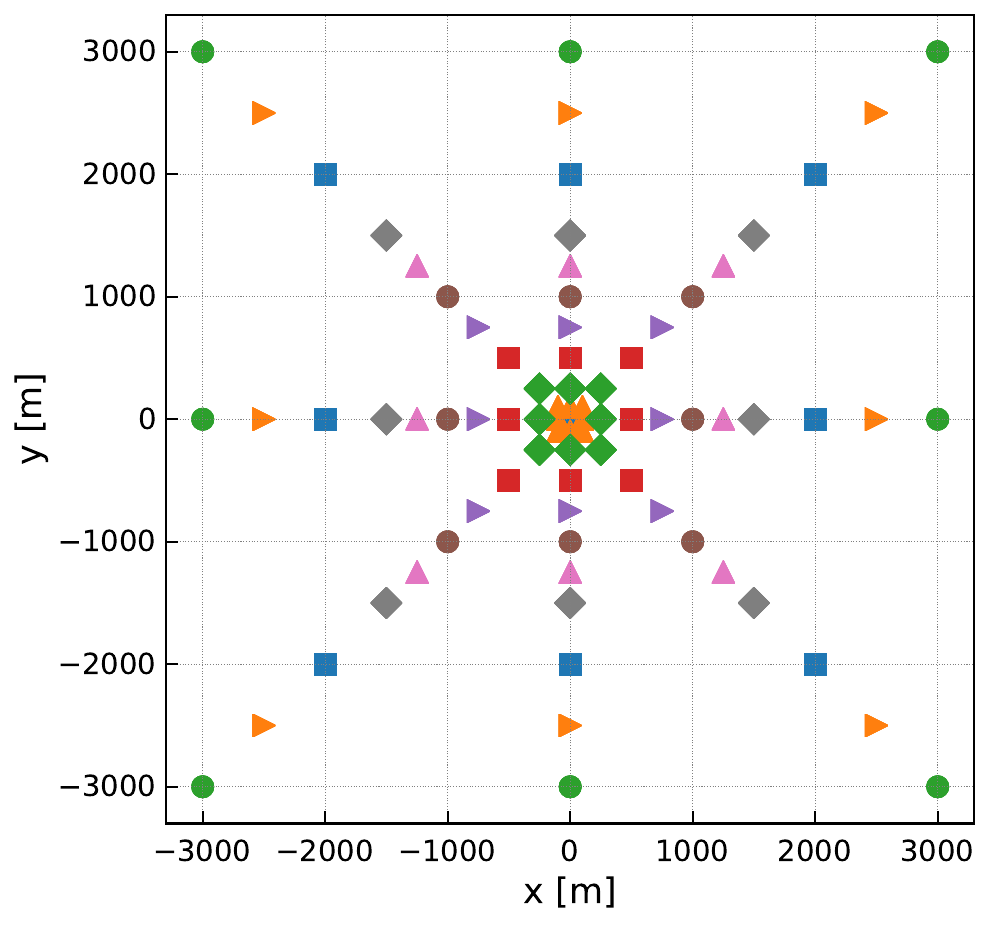}
    \includegraphics[width=0.45\textwidth]{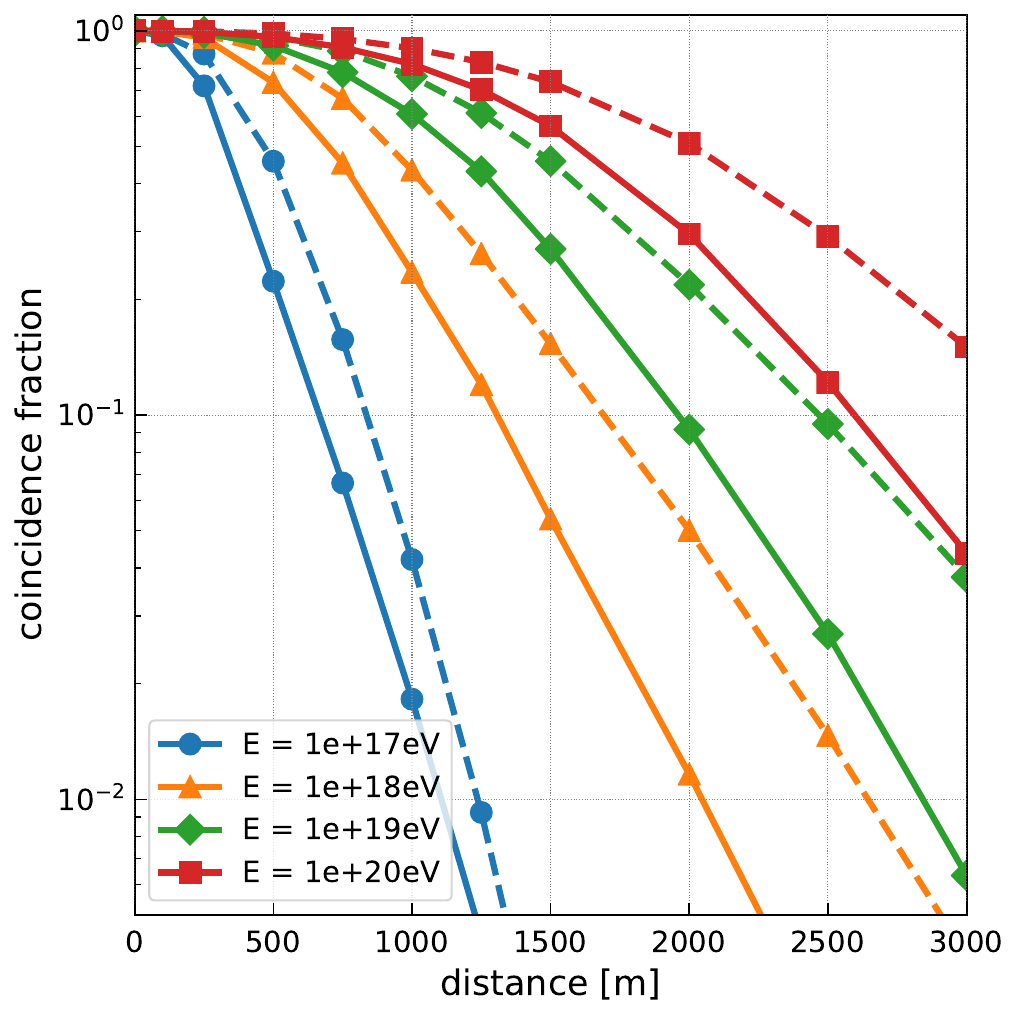}
    \caption{(top) Station layout of example 3 to determine the station coincidence rate. Each color and symbol combination corresponds to the nearest neighbors of one station separation distance. (bottom) The probability to detect the same neutrino in multiple stations as a function of separation between the stations. The different colors/symbols correspond to different neutrino energies. The solid line is the result for a surface detector, the dashed line is the result for a \SI{50}{m} deep detector.}
    \label{fig:example3_layout}
\end{figure}

For every station separation distance we consider the eight nearest stations around the central station as illustrated in Fig.~\ref{fig:example3_layout} on the left. We consider distances ranging from \SI{100}{m} to \SI{3}{km}.

We run the \NuRadioMC\ simulation for event lists of different neutrino energies. The Askaryan signal is filtered from \SI{80}{MHz} - \SI{500}{MHz} and all events are saved that exceed a signal threshold of $1V_\mathrm{RMS}$ for a noise temperature of \SI{300}{K}. 

\subsection{Accessing the results and coincidence fraction}

Part of the HDF5 output file is the maximum amplitude of each channel of each event stored in a two dimensional array. This allows for a quick calculation of the coincidence requirements. We first check if the central station fulfilled the trigger condition which we assume to be a signal above $3V_\mathrm{RMS}$ in any channel. Then, for each simulated distance, we select the channels corresponding to this distance and check if any channel fulfills the trigger condition. The coincidence rate is then given by the ratio of events where both the central station and any of its nearest neighbors triggered, divided by the number of triggers of the central station alone. The result is presented in Fig.~\ref{fig:example3_layout} (right). It shows that the coincidence fraction increases with energy. At a station distance of \SI{1}{km} more than 20\% of the events at $10^{18}$ eV for a surface station (and more than 40\% for a \SI{50}{m} deep station) are detected in at least two stations. This suggests that for a design optimizing on effective volume, stations should be separated further than \SI{1}{km} from each other, or even further when optimizing for the highest energies. An array of surface stations shows in general a smaller coincidence fraction. 

\section{Summary and Outlook}
We have presented \NuRadioMC\ as a versatile framework to simulate different aspects of radio neutrino detectors. \NuRadioMC\ provides a state-of-the-art implementation of the four pillars of a radio neutrino simulation: \emph{event generation, signal generation, signal propagation,} and \emph{detector simulation}. All properties of the simulation chain can be adapted and compared to each other. Following the design goals of flexibility and usability, \NuRadioMC\ combines the knowledge and experience from all previous radio detectors for neutrino and cosmic-rays. We have presented a detailed discussion of many radio emission models and documented an improved time-domain approach using a shower library which provides a realistic treatment of the LPM effect and its random fluctuations. In three comprehensive examples, we have shown how to calculate effective volumes and sensitivities, the efficiency to detect multiple pulses from the same shower (multi-path events), and the coincidence fraction between stations in a large array, depending on the distance between stations. This provides valuable tools for design decisions, depending on the goals one wants to optimize for. Proposed radio neutrino experiments such as RNO, ARIANNA, GRAND, ANITA/PUEO or BEACON \cite{RNO,ARIA,GRAND,BEACON} may soon or already have profited from the capabilities of \NuRadioMC.  

\NuRadioMC\ provides a solid foundation for reliable simulations, but also leaves room for future developments from the radio neutrino community. \NuRadioMC\ is publicly available on github \cite{NuRadioMC_github} and is open to low-threshold further code development from interested parties. As experiments progress and as soon as neutrinos are detected through their radio emission, the areas of prioritized need for development will be indicated by the data. 

\section{Acknowledgements}
This article and \NuRadioMC\ itself would not exist without the constructive spirit of the InIceMC working group of the ARA and ARIANNA collaborations. 

We acknowledge funding from the German research foundation (DFG) under grant GL 914/1-1 (CG) and grant NE 2031/2-1 (DGF, AN, IP, and CW).
JAM is supported by Ministerio de Econom\'\i a, Industria y Competitividad (FPA 2017-85114-P), Xunta de Galicia (ED431C 2017/07), Feder Funds, RENATA Red Nacional Tem\'atica de Astropart\'\i culas (FPA2015-68783-REDT) and Mar\'\i a de Maeztu Unit of Excellence (MDM-2016-0692).  We are grateful to the U.S. National Science Foundation-Office of Polar Programs, the U.S.
National Science Foundation-Physics Division (grant NSF-1607719) and the U.S. Department of Energy (SWB and CP). BAC thanks the National Science Foundation for support through the Graduate Research Fellowship Program Award DGE-1343012. AC acknowledges funding from the NSF CAREER award 28820 and NSF award 49285. We acknowledge Belgian Funds for Scientific Research (FRS-FNRS and FWO) (ST and NvE), the FWO programme for International Research Infrastructure, and funds from the ERC-StG (no. 805486) of the European Research Council (KDdV).

\bibliographystyle{JHEP}

\bibliography{BIB}

\appendix

\section{HDF5 event files structure}
\label{sec:event_files}

The HDF5 files created by the event generator consist of a collection of arrays containing the properties of the neutrinos and other secondary particles (taus, for instance). The array keys and contents are the following:
\begin{itemize}
    \item \emph{azimuths}, the arrival azimuth angles in radians.
    \item \emph{zeniths}, the arrival zenith angles in radians.
    \item \emph{xx, yy}, and \emph{zz}, the $x$, $y$ and $z$ coordinates in meters for the point where the particles interact or decay.
    \item \emph{event\_ids}, the event identification numbers
    \item \emph{n\_interaction}, the interaction number. $1$ indicates a neutrino interaction, $2$ and greater indicates decay or interaction of a lepton created after the neutrino interaction.
    \item \emph{flavors}, neutrino flavors. $12$ for electron neutrino, $14$ for muon neutrino, and $16$ for tau neutrino. Antineutrinos are represented by $-12$, $-14$, and $-16$.
    A value of $15$ indicates a tau lepton. The numbers are following the standard of \cite{PDG}.
    \item \emph{energies}, the particle energies in electronvolts
    \item \emph{interaction\_type}, the interaction type. \emph{'cc'} for charged current, and \emph{'nc'} for neutral current. \emph{'tau\_had', 'tau\_em', 'tau\_mu'} indicate the tau decays into the ha\-dronic, electromagnetic and muonic channels respectively.
    \item \emph{inelasticities}, the inelasticities for the neutrino interactions and the tau decays, that is, the energy fractions taken by the product cascades.
\end{itemize}
In these HDF5 files we also save as attributes the number of events and the characteristics of the fiducial and total simulated volumes, along with maximum and minimum energies and angles for the neutrinos.

\section{NuRadioMC HDF5 output files structure}

NuRadioMC creates as output an HDF5 file with information on the events and on the simulation outcome. The user can choose between saving all the information for all events or only for those that have triggered. The NuRadioMC HDF5 output files contain all the values that can be found in the event files (\ref{sec:event_files}), along with the following additional arrays:
\begin{itemize}
    \item \emph{triggered}, with ones indicating a triggering event and zeroes a non-triggering event.
    \item \emph{weights}, the weights given to each event as a consequence of propagation through the Earth.
    \item \emph{multiple\_triggers}, indicates if the triggering condition has been met individually for each simulated trigger. The first axis of this array gives the event number, and the second the type of trigger.
\end{itemize}
    The rest of the output arrays are stored in several HDF5 groups, each group corresponding to a simulated station. 
    The following arrays (except for the SNRs array) contained within the station group are multidimensional. Their first axis is the event number, and the second one the antenna. Each group for a given station contains:
\begin{itemize}
    \item \emph{SNRs}, the signal to noise ratios for each event defined as the maximum signal amplitude divided by the RMS noise.
     \item \emph{triggered}, with ones indicating a triggering station and zeroes a non-triggering station.
     \item \emph{multiple\_triggers}, indicates if the triggering condition has been met individually for each simulated trigger. The first axis of this array gives the event number, and the second the type of trigger.
    \item \emph{maximum\_amplitudes}, the maximum amplitudes for the voltages of each antenna.
    \item \emph{maximum\_amplitudes\_envelope}, the maximum amplitudes of the voltage envelope of each antenna.
    \item \emph{travel\_distances}, the distances traveled by the rays. There can be up to two, one for each ray-tracing solution. The third axis of the array indicates the ray-tracing solution. The same principle applies to all arrays containing ray-tracing information.
    \item \emph{travel\_times}, the times taken by the rays from emitter to observer.
    \item \emph{ray\_tracing\_C0}, $C_0$ parameters for the ray tracing solutions.
    \item \emph{ray\_tracing\_C1}, $C_1$ parameters for the ray tracing solutions.
    \item \emph{ray\_tracing\_solution\_type}, strings containing the type of ray tracing solutions: direct, reflected, or refracted.
 \end{itemize}   
    The following arrays of the HDF5 group contain three-dimensional vectors, and therefore they have a fourth axis that allows us to find the $x$, $y$, and $z$ components of said vectors.
\begin{itemize}
    \item \emph{launch\_vectors}, the launch vectors for the ray tracing solutions.
    \item \emph{receive\_vectors}, the receive vectors for the ray tracing solutions.
    \item \emph{polarization}, the polarization of the electric field.
\end{itemize}

In the attributes of the output files the names of the simulated triggers (using the string \emph{trigger\_names}) can be found.

\section{Analytic ray tracing}
The analytic ray tracing in \NuRadioMC\ provides a novel and fast solution of the ray-tracing problem. For completeness we provide the full derivation of the analytic solution, the path, the path length and the travel time. 

\subsection{Derivation of analytic solution}
\label{sec:analytic_ray_tracing_derivation}
In this section, we will derive the analytic solution to the ray tracing problem. Fermat's principle states that the optical path of a ray of light travelling between two points is stationary.  Suppose the index of refraction depends on one coordinate in a three-dimensional Cartesian coordinate system:

\begin{equation}
n(x,y,z) = n(z)
\end{equation}

Further, let $dx/dz = \dot{x}$ and $dy/dz = \dot{y}$, so that the metric may be expressed as:

\begin{equation}
ds = \sqrt{dx^2 + dy^2 + dz^2} = dz \sqrt{\dot{x}^2 + \dot{y}^2 + 1}
\end{equation}

The symmetry of $n(z)$ implies that the coordinate system may be rotated such that $\dot{x} = 0$.  Thus the metric becomes

\begin{equation}
ds = dz \sqrt{\dot{y}^2 + 1}
\end{equation}

Inserting this metric into Fermat's Principle gives

\begin{align}
S &= \int_A^B n ds \\
\delta S &= 0 \\
\delta \int_A^B n(z) \sqrt{1+ \dot{y}^2} dz &= 0
\end{align}

Defining $u = \dot{y}$ and applying the Euler-Lagrange equations yields

\begin{equation}
\dot{u} = -\frac{\dot{n}}{n}(u^3+u) \label{eq:main}
\end{equation}

Letting $v = -\ln n$, Eq.~\ref{eq:main} simplifies to 

\begin{equation}
\dot{u} = \dot{v}(u^3+u) \label{eq:main2}
\end{equation}

Noting that $\dot{v} = dv/dz$, and applying the chain rule gives

\begin{equation}
\frac{du}{dz} \frac{dz}{dv} = \frac{du}{dv} = u^3 + u 
\end{equation}

Rearranging and then integrating gives

\begin{align}
\int \frac{du}{u^3 + u} &= \int dv \\
\ln u - \frac{1}{2}\ln(u^2 + 1) &= v + C_0 \label{eq:int_u}
\end{align}

Equation \ref{eq:int_u} may be solved for $dz/dy$ after re-scaling $C_0$:

\begin{equation}
\frac{dz}{dy} = \pm \sqrt{C_0^2 n^2 - 1} \label{eq:main3}
\end{equation}

In the case of South Pole and Moore's Bay glacial ice, it is found that $n(z)$ is described to within a few percent by an exponential function \cite{Barwick_2018} which allows us to proceed further in solving for the ray-path.

\begin{equation}
n(z) = n_{ice} - \Delta_n \exp(z/z_0) \label{eq:n}
\end{equation}

Let $\gamma=\Delta_n \exp(z/z_0)$, which implies

\begin{align}
n(z) &= n_{ice}-\gamma \\
dz &= \gamma^{-1} z_0 d\gamma
\end{align}

Inserting Eq. \ref{eq:n} into Eq. \ref{eq:main3} and integrating, with $b = 2 n_{ice}$ and $c = n_{ice}^2 - C_0^{-2}$:

\begin{equation}
\int \frac{d\gamma}{\gamma (\gamma^2 - b\gamma + c)^{1/2}} = \pm C_0 \left(\frac{y}{z_0} + C_1\right) \label{eq:main4}
\end{equation}

The second integration constant is $C_1$.  Intriguingly, for depths much greater than the scale height ($|z_i| \gg z_0$, $z_i < 0$), the integral in Eq. \ref{eq:main4} has a singularity in the denominator when the ray is initially horizontal.  This is discussed further below. The solution to Eq. \ref{eq:main4} is available in standard tables.  The solution with $y$ as a function of $z$ via $\gamma$ is:

\begin{multline}
y(z) = \pm C_0^{-1} c^{-1/2} z_0 \\ \ln \left( \frac{\gamma}{2 c^{1/2} (\gamma^2 - b\gamma +c)^{1/2} -b\gamma + 2c} \right) \mp z_0 C_1 \label{eq:main5}
\end{multline}

Let the function within the logarithm in Eq. \ref{eq:main5} be $F(\gamma)$:

\begin{equation}
F(\gamma) = \frac{\gamma}{2 c^{1/2} (\gamma^2 - b\gamma +c)^{1/2} -b\gamma + 2c} \label{eq:Fgamma}
\end{equation}

Inserting Eq. \ref{eq:Fgamma} into Eq. \ref{eq:main5}, we recover a function which returns the ray path as a function of depth:

\begin{equation}
y(z) = \pm C_0^{-1} c^{-1/2} z_0 \ln \left( F(\gamma) \right) \mp z_0 C_1 \label{eq:main5b}
\end{equation}

Because the ice model is horizontally symmetric, the constant $C_1$ is set by the choice of origin.  All that remains is to understand the physical meaning of $C_0$.  Let the initial angle with respect to the horizontal be $\theta_i$, which should obey

\begin{align}
\frac{dy}{dz} &= \cot(\theta_i) \\
\frac{dy}{d\gamma} &= z_0 \gamma^{-1} \cot(\theta_i) \label{eq:der}
\end{align}

Given Eq. \ref{eq:main5b}, Eq. \ref{eq:der} may be solved in terms of $F(\gamma)$.  The result is

\begin{equation}
\tan\theta_i = \pm C_0 c^{1/2} \frac{F(\gamma)}{\gamma F'(\gamma)}
\end{equation}

Inserting the definition of $c$ and solving for $C_0$:

\begin{equation}
C_0(\gamma,\theta_i) = \pm n_{ice}^{-1} \left( \frac{\gamma^2 F'^2(\gamma)}{F^2(\gamma)}\tan^2\theta_i + 1 \right)^{1/2} \label{eq:almost}
\end{equation}

The right-hand side of Eq. \ref{eq:almost} resembles a secant function.  Restricting to initial depths much greater than the scale depth ($|z_i| \gg z_0$, $z_i < 0$) causes

\begin{equation}
\frac{\gamma^2 F'^2(\gamma)}{F^2(\gamma)} \to 1 \label{eq:oh}
\end{equation}

If this limit is taken, then Eq. \ref{eq:almost} simplifies:

\begin{equation}
C_0(\gamma,\theta_i) = \pm n_{ice}^{-1} \left(\tan^2\theta_i + 1 \right)^{1/2} = \pm n_{ice}^{-1} \sec\theta_i \label{eq:gotheem}
\end{equation}

$C_0$ is a constant that depends on the boundary conditions, so Eq.~\ref{eq:gotheem} may be inverted:

\begin{equation}
n_{ice} \cos\theta_i = \pm C_0^{-1} \label{eq:finalC0}
\end{equation}

Equation \ref{eq:finalC0} is Snell's Law, because $C_0$ is constant and $\theta_i$ is defined with respect to the horizontal.  Thus, in the limit ($|z_i| \gg z_0$, $z_i < 0$) the singularity in Eq.~\ref{eq:main4} is for $\cos\theta_i = \pm 1$, i.e. horizontal propagation.  Further, in the limit ($|z_i| \gg z_0$, $z_i < 0$) the factor in front of Eq.~\ref{eq:main5b}, $C_0^{-1} c^{-1/2}$, simplifies:

\begin{align}
c &= n_{ice}^2 - C_0^{-2} \\
c^{-1/2} &= \left( n_{ice}^2 - C_0^{-2} \right)^{-1/2} \\
C_0^{-1} c^{-1/2} &= \left( C_0^2 n_{ice}^2 - 1 \right)^{-1/2} \\
C_0^{-1} c^{-1/2} &= \cot(\theta_i)
\end{align}

In the last step, Eq. \ref{eq:main3} has been used.  Thus, the closed form of $y(z)$ is

\begin{equation}
y(z) = \pm z_0 \cot{\theta_i} \ln \left( F(\gamma) \right)
\end{equation}

If the depth $z$ does not satisfy the limit ($|z_i| \gg z_0$, $z_i < 0$), $C_0$ must first be obtained from Eq. \ref{eq:almost}, and then inserted into Eq. \ref{eq:main5b} to obtain the ray-tracing path.

\subsection{Putting the analytic solution into practical usability}
\label{sec:analytic_ray_tracing_fit}
In this section, we demonstrate how to efficiently solve the analytic equations for the ray path derived in \ref{sec:analytic_ray_tracing_derivation}.
Without loss of generality, we can use only the positive solution which corresponds to rays propagating into the positive $y$ direction. Equally, we can only consider rays in the $y-z$ plane. This is because such a start configuration can always be achieved with a simple coordinate transformation. 

In addition, it is sufficient to only compute solution from a deeper to a shallower position without loss of generality by flipping the initial condition. Hence we can always reduce the problem to finding all possible path's between two points 
\begin{multline}
 \vec{x}_1 = (y_1, z_1)^T \, \mathrm{ and }\,  \vec{x}_2 = (y_2, z_2)^T \\ \mathrm{ with }\, y_1 < y_2\, \mathrm{ and }\, z_1 < z_2 \,.
\end{multline}

The analytic solution only describes the ``first part'' of the solution until the \emph{turning point}. This is the position where the ray either hits the surface and is reflected down, or it reaches the point where the propagation direction of the ray becomes horizontal (i.e. into the y direction) due to continuous refraction. This is of course a consequence of the solution being $y(z)$ and not $z(y)$ which is needed to describe the ray path in a single analytic function (because $z(y)$ is not bijective). 

The turning point is the position where the second root of Eq.~\eqref{eq:analytic_solution} becomes undefined, i.e., for
\begin{equation}
 \gamma^2 - b \gamma + c = 0  \Rightarrow \gamma_\mathrm{turn} = \frac{1}{2} b - \sqrt{\frac{b^2}{4} - c} \, .
\end{equation}
The $z_\mathrm{turn}$ position can be calculated from $\gamma_\mathrm{turn}$. If $z_\mathrm{turn}$ is positive, the turning points is above the surface. Hence, the ray is reflected off the surface and $z_\mathrm{turn}$ is set to zero. Then, $y_\mathrm{turn}$ can be calculated by inserting $z_\mathrm{turn}$ into Eq.~\eqref{eq:analytic_solution}. 

Hence, from an implementation perspective, we have two distinct cases: either we have a direct ray ($y_2 < y_\mathrm{turn}$) or we have a reflected or refracted ray ($y_2 > y_\mathrm{turn}$)

\subsection{Determination of free parameters}
Now, we present how to determine the two free parameters $C_0$ and $C_1$ in a fast and robust way from the initial condition that the ray path goes through the points $\vec{x}_1$ and $\vec{x}_2$. The parameter $C_1$ is given by
\begin{equation}
 C_1 = y_1 - y(z_1, C_0 = C'_0, C_1=0) \,
\end{equation}
with $y()$ being Eq.~\eqref{eq:analytic_solution} evaluated for $C_0 = C'_0$ and $C_1 = 0$. 

The parameter $C_0$ needs to be determined numerically by minimizing the following objective function:
\begin{equation}
 \chi^2 =  \left(y_2 - y'(z_2, C_0, C_1)\right)^2 \, .
\end{equation}

\begin{figure*}[t]
 \centering
 \includegraphics[width=1\textwidth]{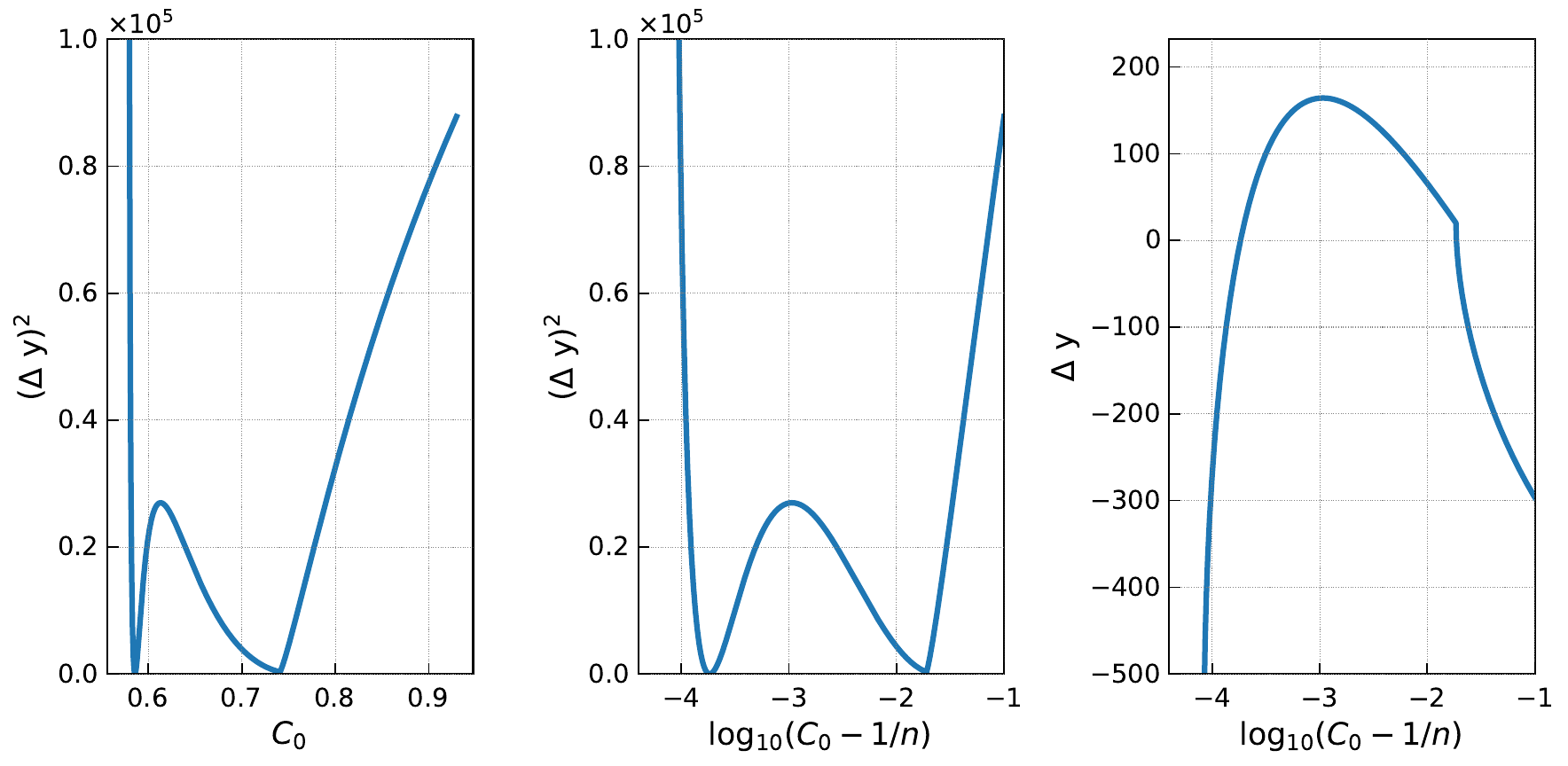}
 \caption{Example of a typical objective function as a function of $C_0$ (left) and $\log_{10}(C_0 - 1/n)$ (center). Displacement in $y$ as used for the determination of the second solution via the root finding algorithm (right).}
 \label{fig:chi2coord}
\end{figure*}


As Eq.~\eqref{eq:analytic_solution} describes only half of the solution, we first check if $\vec{x}_2$ is before or after the turning point. It is after the turning point if $y_\mathrm{turn} < y_2$. Then the following coordinate transformation is performed.
\begin{equation}
 y'(z_2, C_0, C_1) = 2 \, y_\mathrm{turn} - y(z_2, C_0, C_1) \, .
\end{equation}
To increase the numerical stability of the minimizer it is useful to perform the following coordinate transformation
\begin{equation}
 D = \ln(C_0 - 1/n_\mathrm{ice}) \, .
\end{equation}
Then Eq.~\eqref{eq:analytic_solution} is defined for all values of $D$. 

For typical geometries not just one but two solutions are present. Once one solution is found, the second solution can be determined fast and efficiently using the Brent root finding algorithm \cite{Brent1973}, and using the displacement in $y$ at position $\vec{x}_2$ as objective function (cf. Fig.~\ref{fig:chi2coord} right). Utilization of Brent's algorithm is possible because for a second solution to exists, $\Delta y$ needs to change sign in one of the open intervals $(-\infty, C_0^1)$ and  $(C_0^1, \infty)$, where $C_0^1$ is the first solution.

 
 \subsection{Derivative of analytic ray tracing path}
 \label{sec:analytic_ray_tracing_derivative}
 The derivative of the analytic ray tracing solution is given by
 \begin{multline}
\frac{d y(z)}{d z} =  \\ \left( -\sqrt {c}{{\rm e}^{{\frac {z}{{\it z_0}}}}}b{\it \Delta_n}+2
\,\sqrt {-b{\it \Delta_n}\,{{\rm e}^{{\frac {z}{{\it z_0}}}}}+{{\it 
\Delta_n}}^{2}{{\rm e}^{2\,{\frac {z}{{\it z_0}}}}}+c}\, c+2\,{c}^{3/2}
 \right)\\ \times
 \left( 2\,\sqrt {c}\sqrt {-b{\it \Delta_n}\,{{\rm e}^{{
\frac {z}{{\it z_0}}}}}+{{\it \Delta_n}}^{2}{{\rm e}^{2\,{\frac {z}{{
\it z_0}}}}}+c}-b{\it \Delta_n}\,{{\rm e}^{{\frac {z}{{\it z_0}}}}}+
2\,c \right) ^{-1}\\  \times 
{\frac {1}{\sqrt {-b{\it \Delta_n}\,{{\rm e}^{{
\frac {z}{{\it z_0}}}}}+{{\it \Delta_n}}^{2}{{\rm e}^{2\,{\frac {z}{{
\it z_0}}}}}+c}}}{\frac {1}{\sqrt {{{\it C_0}}^{2}{{\it n_\mathrm{ice}}}^{2}
-1}}} \, .
\end{multline}

\subsection{Analytic solution of path length and travel time}
\label{appendix:analyticT}
In this section, the analytic solution of the path length and travel time for an exponential index-of-refraction profile is derived. 

To find the path(s) between two given points in the ice, $(r_0, z_0)$ and $(r_1, z_1)$, we need to find the launch angle(s) $\theta_0$ of the ray(s). The radial coordinate $r$ is equivalent to the $y$ coordinate used in the previous sections, since we are restricted to the vertical plane where the wave propagates.
Given the launch angle $\theta_0$ then we can find $\theta$ as a function of $z$ using Snell's Law:

\begin{equation} \label{eq:snell}
n(z) \sin(\theta(z)) = n(z_0) \sin(\theta_0)
\end{equation}

\begin{equation} \label{eq:theta}
\theta(z) = \arcsin\left( \frac{n(z_0) \sin(\theta_0)}{n(z)} \right)
\end{equation}

Since we know the radial distance between our starting and ending points, we can calculate the launch angle by first working out the radial distance integral as a function of launch angle, and then inverting it.


\begin{equation*}
\frac{dr}{dz} = \frac{dr}{ds} \frac{ds}{dz} = \tan(\theta)
\end{equation*}

\begin{equation*}
\int_{r_0}^{r_1}\ dr = \int_{z_0}^{z_1} \tan(\theta)\ dz
\end{equation*}

And then using equation \ref{eq:theta}, this becomes

\begin{equation} \label{eq:rintgeneral}
r_1 - r_0 = \int_{z_0}^{z_1} \tan\left( \arcsin\left( \frac{n(z_0) \sin(\theta_0)}{n(z)} \right) \right)\ dz
\end{equation}

To calculate the launch angle(s) for ray(s) between our two points, solve this equation for $\theta_0$.
While we will continue solving this problem in generality for any $n(z)$ now, in a following section we will simplify the answer for a specific ice model.

Once we know the launch angle of our path we have all we need to calculate its properties.
The total path length can be calculated by integrating $\frac{dz}{ds}$:

\begin{align} \label{eq:sintgeneral}
s = &\int_{z_0}^{z_1} \frac{1}{\cos(\theta)}\ dz\\
= &\int_{z_0}^{z_1} \sec\left( \arcsin\left( \frac{n(z_0) \sin(\theta_0)}{n(z)} \right) \right)\ dz
\end{align}

The time of flight $t$ along the path can be calculated by combining $\frac{dz}{ds}$ with the following differential equation for the time of flight (where $c$ is the speed of light):

\begin{equation} \label{eq:dtds}
\frac{dt}{ds} = \frac{n(z)}{c}
\end{equation}
Which then gives
\begin{equation*}
\frac{dt}{dz} = \frac{dt}{ds} \frac{ds}{dz} = \frac{n(z)}{c} \frac{1}{\cos(\theta)}
\end{equation*}

\begin{align} \label{eq:tintgeneral}
t =& \int_{z_0}^{z_1} \frac{n(z)}{c} \frac{1}{\cos(\theta)}\ dz \\
= &\frac{1}{c} \int_{z_0}^{z_1} n(z) \sec\left( \arcsin\left( \frac{n(z_0) \sin(\theta_0)}{n(z)} \right) \right)\ dz
\end{align}

For an exponential index-of-refraction profile of the form
\begin{equation} \label{eq:icemodel}
n(z) = n_\textrm{ice} - \Delta_n e^{z/z_0}
\end{equation}
we can finish the calculations.
We will use a few substitutions to make our equations clearer.
The substitutions are as follows, where $n(z)$ is as above, $z_0$ is the starting depth, and $\theta_0$ is the launch angle:
\begin{equation} \label{eq:substitutions}
\begin{split}
\beta &= n(z_0) \sin(\theta_0)					\\
\alpha &= n_\textrm{ice}^2 - \beta^2						\\
\gamma &= n(z)^2 - \beta^2						\\
\ell_1 &= n_\textrm{ice} n(z) - \beta^2 + \sqrt{\alpha \gamma}	\\
\ell_2 &= n(z) + \sqrt{\gamma}
\end{split}
\end{equation}

Plugging in our ice model, the radial distance integral in equation \ref{eq:rintgeneral} becomes\footnote{Equations \ref{eq:substitutions}, \ref{eq:rint}, \ref{eq:sint}, and \ref{eq:tint} are slightly different from those originally published. The equations shown here are (up to an integration constant) equivalent to those published previously, but do not suffer from numerical instability at larger depths $z$. Please see \cite{Bouma:2025qvh} for a more detailed discussion. }
\begin{equation} \label{eq:rint}
r_1 - r_0 = \left.\frac{\beta}{\sqrt{\alpha}} \left( z -
z_0 \log \left(\ell_1\right) \right) \right|_{z_0}^{z_1}
\end{equation}
after equation \ref{eq:substitutions}'s substitutions. 
Solving this equation for the launch angle is an alternative approach to find the ray tracing path. 
Unfortunately, since the launch angle appears in so many places ($\alpha$, $\beta$, and $\ell_1$), this equation is not invertible and so cannot be directly solved for $\theta_0$.
As a result, root-finding algorithms will need to be used to calculate the launch angle(s) for the ray(s) between $(r_0, z_0)$ and $(r_1, z_1)$. In the \NuRadioMC\ code, we calculate the ray paths using the approach of Sec.~\ref{sec:analytic_ray_tracing_fit} and just calculate the launch angle from the parameter $C_0$ of the analytic ray-tracing path. 

Plugging in our ice model and substituting according to equation \ref{eq:substitutions}, the path length (equation \ref{eq:sintgeneral}) becomes

\begin{equation} \label{eq:sint}
s = \left.\frac{n_\textrm{ice}}{\sqrt{\alpha}} \left(z - z_0 \log(\ell_1)\right) + z_0 \log(\ell_2) \right|_{z_0}^{z_1}
\end{equation}

By the same process, the time of flight (equation \ref{eq:tintgeneral}) becomes

\begin{equation} \label{eq:tint}
t = \left.\frac{1}{c} \left(z_0 \left(\sqrt{\gamma} + n_\textrm{ice} \log(\ell_2) - \log(\ell_1) \frac{n_\textrm{ice}^2}{\sqrt{\alpha}}\right)
+ z \frac{n_\textrm{ice}^2}{\sqrt{\alpha}}\right) \right|_{z_0}^{z_1}
\end{equation}


Note that these integrals are specifically for a direct path.
For an indirect path, the bounds must be changed to reflect the fact that the path goes up to $z_\text{turn}$ before coming back down to $z_1$.

\subsection{Derivation of focusing correction}
\label{appendix:focusing}

\begin{figure}
    \centering
    \begin{tikzpicture}[line width=1.5pt]
        \node (i1) at (0,0) {};
        \node (iup) at (0,2.75) {};
        \node (o1) at (6,1) {};
        \node (o2) at (6,2.5) {};
        \node (rperp) at ($(i1)!(o1)!(o2)$) {};

        \path
        (i1.center) edge [edge label=$R$] (o2.center)
        (i1.center) edge (o1.center)
        (o1.center) edge[blue, edge label=$a_\theta$] (rperp.center)
        (o1.center) edge [red, edge label=$\Delta z$, swap] (o2.center)
        (i1.center) edge[dotted, gray] (iup.center)

        pic [draw, "$\theta$", thin] {angle = o2--i1--iup}
        pic [draw, "$\Delta\theta$", thin, angle radius=2cm, angle eccentricity=0.8] {angle = o1--i1--o2}
        pic [draw, "$\theta'$", thin] {angle = i1--o2--o1}
        ;
    \end{tikzpicture}
    \caption{Sketch of geometry for focusing correction.}
    \label{fig:focusing_correction}
\end{figure}
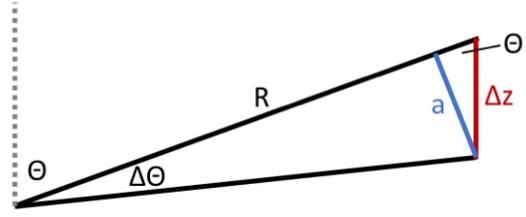
Here, we derive how ray density per unit area changes. The geometry in case of straight-line propagation is depicted in Fig.~\ref{fig:focusing_correction}. We read off that $a_\theta = R \, \sin\Delta\theta$. In the limit of $\Delta \theta << 1$ we get $a_\theta = R \, \Delta\theta$. 
If instead we express this as a function of the vertical displacement $\Delta z$, we get $a_\theta = \Delta z \sin \theta'$. Taking the limit $\Delta z \rightarrow 0$, we obtain:
\begin{equation}
    a_\theta = \frac{\mathrm{d} z}{\mathrm{d} \theta} \sin \theta' \, \mathrm{d} \theta, 
\end{equation}
where only for straight-line propagation as sketched in Fig.~\ref{fig:focusing_correction} we have $\theta'=\theta$.
In the case of a refractive index $n(z)$ that depends only on the depth $z$, there is no ray bending in the $\phi$-direction. The width in this direction is therefore a function only of the horizontal distance $r$:
\begin{equation}
    a_\phi = r \mathrm{d}\phi.
\end{equation}
Note that in the case of straight-line propagation, we have $r = R \sin\theta$. 
Thus, we obtain the familiar formula for the area $\mathrm{d} A$ perpendicular to a ray in the case of no ray bending:
\begin{align}
    \mathrm{d}A &= a_\theta \times a_\phi = R \mathrm{d}\theta \times R \sin\theta \mathrm{d}\phi \, ,
\end{align}
which generalizes to the case of ray bending in the $\theta$-direction only to:
\begin{align}
    \mathrm{d}A &= \frac{\mathrm{d}z}{\mathrm{d}\theta} \sin \theta' \mathrm{d}\theta \times r \mathrm{d}\phi \, .
\end{align}

\section{FFT normalization in \NuRadioMC}
\label{sec:FFT}
In \NuRadioMC\ we use a real fast Fourier transform (rFFT) as it only deals with real valued signals in the time-domain. Furthermore, we assume that the number of samples in the time domain is even. Then, $n_t$ bins (with real values) in the time domain correspond to $n_f = n_t / 2 + 1$ bins (with complex values) in the frequency domain where the first bin is the zero frequency component. This is because we exploit the symmetry between negative and positive frequencies for real valued input and only compute the positive frequency components. 

The rFFT is normalized such that Parseval's theorem holds without any additional normalization factor, i.e., 
\begin{equation}
 \sum\limits_{m=0}^{n_t-1} x_m^2 \cdot \Delta t = \sum\limits_{k=0}^{n_t/2} \tilde{X}_k^2 \cdot \Delta f\, .
 \label{eq:parseval}
 \end{equation}
 where $x_m$ are the time domain samples of the signal, $\tilde{X}_k$ are the frequency domain samples, and $\Delta t, \Delta f$ are the bin widths in the time and frequency domains, respectively. In the case of electric fields, the dimensions of $x_m$ and $\tilde{X}_k$ are voltage/length and voltage/(length $\times$ frequency), respectively\footnote{These dimensions, and the updated form of Eq.~\eqref{eq:parseval}, reflect the FFT implementation since version 1.1.0.}.

This means that the energy fluence, i.e., the time integral over the pulse amplitudes, calculated in the frequency domain and in the time domain give the same results which is a useful physical property. Then, the rFFT and inverse rFFT is defined as 
\begin{equation}
 \tilde{X}_k = \sqrt{2} \Delta t \times \sum\limits_{m=0}^{n_t-1} x_m \exp\left(-2\pi i \frac{mk}{n_t}\right) \,,
 \end{equation}
  for $k = 0, ..., n_t/2$, and
 \begin{equation}
 x_m = \frac{1}{\sqrt{2} \Delta t n_t} \times \sum\limits_{k=-n_t/2+1}^{n_t/2} \tilde{X}_k \exp\left(2\pi i \frac{mk}{n_t}\right) \,,
\end{equation}
for $m = 0, ..., n_t$, where $\tilde{X}_{-k} = \tilde{X}_{k}^\dagger$ because the trace in the time domain is purely real.

The additional factor of $\sqrt{2}$ with respect to the standard normalization was added to compensate for the negative frequencies that we did not compute, so that Eq.~\eqref{eq:parseval} holds\footnote{Note that, for an even trace length, this convention double-counts the zero- and Nyquist-frequency components on the right-hand side of Eq.~\eqref{eq:parseval}, so that the equality only holds approximately. As these two components should generally be small, the difference is usually negligible.}.

\subsection{Relation to a continuous Fourier transform}
In literature, one also finds the continuous Fourier transform with different conventions for the normalization. One typical choice is to define the Fourier transform as
\begin{equation}
 \tilde{E}(\nu) = \int\limits_{-\infty}^\infty dt \exp\left(i 2 \pi \nu t\right) E(t)
 \label{eq:contiuousforward}
\end{equation}
and 
\begin{equation}
  E(t) = \int\limits_{-\infty}^\infty d\nu \exp\left(-i 2 \pi \nu t\right) \tilde{E}(\nu) \, .
  \label{eq:contiuousbackward}
\end{equation}
If the signal in the time domain has units \si{V/m} the units in the frequency domain become \si{V/m/Hz}. A common task is to transform a frequency-domain parameterization of the Askaryan signal into the time domain via a discrete Fourier transform. For the definition of Eq.~\eqref{eq:contiuousforward}, the corresponding discrete inverse transform is
\begin{align}
 x_m = &\frac{1}{n_t} \times 2\sum\limits_{k=0}^{n_t/2} \tilde{X}_k / \Delta t \exp\left(2\pi i \frac{mk}{n_t}\right)\\ = &2\sum\limits_{k=0}^{n_t/2} \tilde{X}_k \Delta f \exp\left(2\pi i \frac{mk}{n_t}\right) \, 
 \label{eq:backwardfft}
\end{align}
where we exploit the relation $\Delta t = 1/(n_t \Delta f)$ of a discrete Fourier transform. The additional factor of $2$ was added because we only sum over the positive frequencies here. This factor of $2$ is already part of real FFT packages such as \emph{numpy.fft} and does not need to be taken into account by the user (see Sec.~\ref{sec:implementation} for details). 

\subsection{Adjustments to different normalizations}
All publications of a frequency-domain parameterization of the Askaryan signal that is based on the \emph{ZHS} model use an unusual normalization of the continuous Fourier transform where an additional factor of $2$ is added to the forward transform (Eq.~\ref{eq:contiuousforward}), and correspondingly a factor of $1/2$ in the backward transform (Eq.~\ref{eq:contiuousbackward}) (see e.g. \cite{Alvarez2009}). Therefore, Eq.~\eqref{eq:backwardfft} needs an additional factor of $1/2$ if a ZHS parameterization is used.

\subsection{Implementation details}
\label{sec:implementation}
Most parts of the code use the numpy real fft routines. The default normalization has the direct transforms unscaled and the inverse transforms are scaled by $1/n_t$. Hence, a analytic parametrization of the amplitudes in the frequency domain $A(\nu)$ with units \si{V/m/Hz} can be transformed into the time domain via

\begin{minted}{python}
import numpy as np
n = 2**12  # number of bins in time domain
dt = 0.5 * units.ns  # bin width in time domain
ff = np.fft.rfftfreq(n, dt)  
    # get array of frequencies
trace = np.fft.irfft(A(ff) / dt)
\end{minted}

If $A(\nu)$ is a parametrization from a ZHS paper, we get the correct time domain representation via 
\begin{minted}{python}
trace = 0.5 * np.fft.irfft(A(ff) / dt) 
    # additional factor of 2 due to 
    # ZHS Fourier transform normalization
\end{minted}

All other Fourier transforms are normalized such that Eq.~\eqref{eq:parseval} is satisfied which is achieved with numpy via:
\begin{minted}{python}
 def time2freq(trace, sampling_rate):
    """
    performs forward FFT with correct 
    normalization that conserves the power
    """
    return np.fft.rfft(trace, 
        axis=-1) / sampling_rate * 2 ** 0.5  
    # an additional sqrt(2) is added because 
    # negative frequencies are omitted.


def freq2time(spectrum, sampling_rate):
    """
    performs backward FFT with correct 
    normalization that conserves the power
    """
    return np.fft.irfft(spectrum, 
        axis=-1) * sampling_rate / 2 ** 0.5
\end{minted}

\begin{listing*}[ht]
\begin{minted}[
%frame=lines,
%framesep=2mm,
%baselinestretch=1.2,
fontsize=\footnotesize,
linenos
]{python}
def get_time_trace(energy, theta, N, dt, shower_type, n_index, R, model, 
                   interp_factor=None, interp_factor2=None,
                   same_shower=False, **kwargs):
    """
    returns the Askaryan pulse in the time domain of the eTheta component

    We implement only the time-domain solution and obtain the frequency spectrum
    via FFT (with the standard normalization of NuRadioMC). This approach assures
    that the units are interpreted correctly. In the time domain, the amplitudes
    are well defined and not details about fourier transform normalizations needs
    to be known by the user.

    Parameters
    ----------
    energy : float
        energy of the shower
    theta: float
        viewangle: angle between shower axis (neutrino direction) and the line
        of sight between interaction and detector
    N : int
        number of samples in the time domain
    dt: float
        time bin width, i.e. the inverse of the sampling rate
    shower_type: string (default "HAD")
        type of shower, either "HAD" (hadronic), "EM" (electromagnetic) or 
        "TAU" (tau lepton induced), note that TAU showers 
        are currently only implemented in the ARZ2019 model
    n_index: float
        index of refraction
    R: float
        distance from vertex to observer
    model: string
        specifies the signal model
        * ZHS1992: the original ZHS parametrization from E. Zas, ...
        * Alvarez2000: parameterization based on ZHS mainly based on J. Alvarez-...
        * Alvarez2009: parameterization based on ZHS from J. Alvarez-...
        * HCRB2017: analytic model from J. Hanson, A. Connolly ...
        * ARZ2019 semi MC time domain model
    interp_factor: float or None
        controls the interpolation of the charge-excess profiles in the ARZ model
    interp_Factor2: float or None
        controls the second interpolation of the charge-excess profiles in the ARZ model
    same_shower: bool (default False)
        controls the random behviour of picking a shower from the library in the ARZ model,
        see description there for more details    

    Returns
    -------
    time trace: array
        the amplitudes for each time bin

    """
\end{minted}
\caption{Signature of the signal generation interface. NuRadioMC provides a uniform interface in form of simple function to all implemented Askaryan modules. This allows to use the Askaryan modules outside of a NuRadioMC simulation and is a well tested resource/reference implementation for the radio community.}
\label{lst:signalgeneration}
\end{listing*}

\section{Detector simulation}
\label{sec:detector_sim}

The code snippet in List.~\ref{lst:detectorsimulation} shows a typical detector simulation. With just a few lines of code, we can calculate the antenna response, downsample the time trace to the detector sampling rate, bandpass filter the signal and simulate a high/low trigger with a 2 out of 4 antennas coincidence logic. 

\begin{listing*}[ht]
\begin{minted}[
%frame=lines,
%framesep=2mm,
%baselinestretch=1.2,
fontsize=\footnotesize,
linenos
]{python}
class mySimulation(simulation.simulation):
    def _detector_simulation(self):
        # 1st convolve efield with antenna pattern
        efieldToVoltageConverterPerChannel.run(self._evt, self._station, self._det)  
        # downsample trace back to detector sampling rate
        channelResampler.run(self._evt, self._station, self._det, sampling_rate=1. / self._dt)
        # bandpass filter the signal
        channelBandPassFilter.run(self._evt, self._station, self._det,
                                  passband=[80 * units.MHz, 500 * units.GHz],
                                  filter_type='butter', order=2)
        # run a high/low trigger on the 4 downward pointing LPDAs
        triggerSimulatorHighLow.run(self._evt, self._station, self._det,
                                    threshold_high=4 * self._Vrms,
                                    threshold_low=-4 * self._Vrms,
                                    coinc_window=40 * units.ns
                                    triggered_channels=[0, 1, 2, 3],  # select the LPDA channels
                                    number_concidences=2,  # 2/4 majority logic
                                    trigger_name='LPDA_2of4_4sigma')
\end{minted}
\caption{Example of performing a detector simulation using NuRadioReco.}
\label{lst:detectorsimulation}
\end{listing*}

\end{document}